\documentclass[a4paper,12pt]{article}

\usepackage[english]{babel}

\usepackage[utf8x]{inputenc}
\usepackage[T1]{fontenc}
\usepackage{ucs}
\usepackage{textcomp}
\usepackage{microtype}
\usepackage{lmodern} 
\usepackage{xspace}

\usepackage{stmaryrd} 
\usepackage{amsmath}
\usepackage{amsfonts}
\usepackage{amssymb}
\usepackage{mathrsfs}
\usepackage{tensor} 

\usepackage{ifthen} 
\usepackage{siunitx} 
\usepackage{import} 

\usepackage{geometry}
\usepackage{fancyhdr} 
\usepackage{hyperref}
\usepackage{cleveref}
\usepackage{url}
\usepackage[table]{xcolor}

\usepackage{graphicx}
\usepackage[small,labelfont={bf},labelsep=period]{caption}
\usepackage{authblk} 

\usepackage{array}
\usepackage{empheq} 

\usepackage[round]{natbib}
\usepackage{doi} 

\usepackage{titlesec} 
\titleformat*{\section}{\bfseries}
\titleformat*{\subsection}{\itshape}
\titleformat{\subsubsection}[runin]{\itshape}
{\thesubsubsection}{.5em}{}[.]

\newcommand\ph{\ensuremath{\varphi}}

\newcommand{\cst}{\mathrm{cst}}

\newcommand\define{\equiv}

\newcommand\vect[1]{\boldsymbol{#1}}
\newcommand\mat[1]{\boldsymbol{#1}}

\newcommand\ex[1]{\mathrm{e}^{#1}}
\newcommand\ii{\mathrm{i}}

\renewcommand\Re{\mathrm{Re}}
\renewcommand\Im{\mathrm{Im}}

\newcommand{\tr}{\mathrm{tr}}

\newcommand\e[1]{_{\mathrm{#1}}}
\newcommand\h[1]{^{\mathrm{#1}}}

\newcommand{\dd}{\mathrm{d}}

\newcommand{\pd}[3][]{\frac{\partial^{#1} #2}{\partial {#3}^{#1}}}
\newcommand{\ddf}[3][]{\frac{\dd^{#1} #2}{\dd {#3}^{#1}}}

\newcommand{\delimiters}[4][]{
\ifthenelse{ \equal{#1}{1} }{  #2 #3 #4  }
					{ \ifthenelse{\equal{#1}{2}}{ \big#2 #3 \big#4 }
						{ \ifthenelse{\equal{#1}{3}}{ \Big#2 #3 \Big#4 }
							{ \ifthenelse{\equal{#1}{4}}{ \bigg#2 #3 \bigg#4 }
								{ \ifthenelse{\equal{#1}{5}}{ \Bigg#2 #3 \Bigg#4 }
									{ \left#2 #3 \right#4 }
								}
							}
						}
					}
													}
													
\newcommand{\pa}[2][]{\delimiters[#1]{(}{#2}{)}}
\newcommand{\pac}[2][]{\delimiters[#1]{[}{#2}{]}}

\definecolor{blue4}{RGB}{0,0,143}
\definecolor{red4}{RGB}{143,0,0}
\definecolor{orange}{RGB}{255,128,0}
\definecolor{darkcyan}{RGB}{0,128,128}
\definecolor{olive}{RGB}{0,128,0}
\definecolor{purple}{RGB}{128,0,128}
\definecolor{cyan2}{RGB}{0,255,255}
\definecolor{fushia}{RGB}{255,0,255}
\definecolor{mygray}{gray}{0.5}
\definecolor{lightgray}{gray}{0.85}

\newcommand{\jacobi}{\mathcal{D}}
\newcommand{\amplification}{\mathcal{A}}
\newcommand{\wronski}{\mathcal{W}}
\newcommand{\hessian}{H}
\newcommand{\tdmatrix}{\mathcal{T}}
\newcommand{\geo}[1]{\mathscr{#1}}

\title{\bfseries Gravitational lenses in arbitrary space-times}

\author[1]{Pierre Fleury\footnote{\href{mailto:pierre.fleury@uam.es}{pierre.fleury@uam.es}}}
\author[2]{Julien Larena\footnote{\href{mailto:julien.larena@uct.ac.za}{julien.larena@uct.ac.za}}}
\author[3,4]{Jean-Philippe Uzan\footnote{\href{mailto:uzan@iap.fr}{uzan@iap.fr}}}

\affil[1]{Instituto de F\'isica Te\'orica UAM-CSIC,
Universidad Aut\'{o}noma de Madrid,\par
Cantoblanco, 28049 Madrid, Spain}
\affil[2]{Department of Mathematics and Applied Mathematics,
University of Cape Town,\par
Rondebosch 7701, South Africa}
\affil[3]{Institut d'Astrophysique de Paris, CNRS UMR 7095,
Sorbonne Universit\'es, \par
98 bis Boulevard Arago, 75014 Paris, France}
\affil[4]{Institut Lagrange de Paris, Sorbonne Universit\'es,\par
98 bis, Boulevard Arago, 75014 Paris, France}

\date{\today}

\begin{document}

\maketitle

\thispagestyle{fancy}
\renewcommand{\headrulewidth}{0pt}
\rhead{{\footnotesize IFT-UAM/CSIC-20-155}}
\cfoot{\thepage}

\begin{abstract}
The precision reached by current and forthcoming strong-lensing observations requires to accurately model various perturbations to the main deflector. Hitherto, theoretical models have been developed to account for either cosmological line-of-sight perturbations, or isolated secondary lenses via the multi-plane lensing framework.
This article proposes a general formalism to describe multiple lenses within an arbitrary space-time background. The lens equation, and the expressions of the amplification and time delays, are rigorously derived in that framework.
Our results may be applied to a wide range of set-ups, from strong lensing in anisotropic cosmologies, to line-of-sight perturbations beyond the tidal regime.
\end{abstract}

\newpage
\pagestyle{fancy}
\fancyhead[L]{\itshape Gravitational lenses in arbitrary space-times}
\fancyhead[R]{\thepage}
\fancyfoot[L,C,R]{}
\renewcommand{\headheight}{15pt}
\renewcommand{\headrulewidth}{0pt}
\renewcommand{\footrulewidth}{0pt}

\setcounter{tocdepth}{2}
\tableofcontents

\bigskip

\hrule

\section{Introduction}
\label{sec:introduction}

Gravitational lensing~\citep{1992grle.book.....S} has become a key probe of the content, structure, and physical laws of our Universe. While weak lensing teaches us about the distribution of matter on cosmic scales~\citep[e.g.][]{Asgari:2020wuj, Gatti:2019clj}, strong lensing lies amongst the best tools to measure today's cosmic expansion rate~$H_0$~\citep{1964MNRAS.128..307R, 2020MNRAS.498.1420W}, and encapsulates valuable information on the small-scale distribution and nature of the dark matter~\citep{2018MNRAS.475.5424D, Diego:2017drh, Harvey:2019bco, Blum:2020mgu}.

The quality of current imaging and data analysis requires an equally elaborate theoretical modelling of the strong lenses. That means complex models for the mass distribution of the main deflector, responsible for, e.g., the formation of multiple images of the same source, but also of the secondary deflectors which may perturb the effect of the main lens~\citep{1997ApJ...482..604K,2018MNRAS.477.5657T,Sengul:2020yya,Li:2020fpq}. Such perturbations are referred to as line-of-sight effects.

Two distinct frameworks have been developed to model line-of-sight perturbations in strong gravitational lensing. The first approach~\citep{1987ApJ...316...52K,1994A&A...287..349S,1996ApJ...468...17B,Schneider:1997bq,Birrer:2016xku} consists in treating secondary deflectors in the tidal regime, i.e., adding external convergence and shear to the main lens model. This technique is well suited to describe cosmological perturbations which would add to the effect of a lens; it has been successfully applied to measuring cosmic shear with Einstein rings by \citet{Kuhn:2020wpy}. On the contrary, the second approach consists in modelling all the secondary deflectors as thin lenses. This formalism, called multi-plane lensing~\citep{1986ApJ...310..568B}, is thus adapted to the description of isolated perturbers within an otherwise empty Universe, or within an ideal Friedmann-Lemaître-Robertson-Walker (FLRW) cosmology. The applicability of the original multi-plane formalism was then extended by the introduction of tidal planes~\citep{McCully:2013fga}, voids~\citep{McCully:2016yfe}, or cosmological perturbations~\citep{Schneider:2014vka}; however, it has never been considered in full generality.

The present article proposes to fill this gap with an ultimate multi-plane formalism, where one or several lenses with arbitrary velocities may be placed in any smooth space-time background. This formalism encapsulates all the key lensing observables into a single versatile language. Our results may be applied to a wide range of set-ups; three specific examples will be provided in this article: (1) one lens with cosmological perturbations; (2) one lens in anisotropic cosmology, which recently attracted renewed attention~\citep{Akrami:2019bkn,Migkas:2020fza, 2021ApJ...908L..51S, Fosalba:2020gls}; and (3) multiple lenses in an under-dense Universe. Furthermore, this work establishes the necessary tools to accurately describe line-of-sight corrections in strong lensing beyond the standard convergence and shear~\citep{Fleury:2018odh, FLU21}.

The article is organised as follows. We start in \cref{sec:preliminary_discussion} with a discussion on the nature of the gravitational fields encountered by light beams, thereby defining the dichotomy between reference space-time and lenses. In \cref{sec:one_lens} we consider the case where a single lens is placed within an arbitrary reference space-time, before moving to the general $N$-lens case in \cref{sec:N_lenses}. We conclude in \cref{sec:conclusion}.

We adopt units for which the speed of light is unity. Bold symbols ($\vect{\alpha}, \vect{\beta}, \ldots$) indicate two-dimensional Euclidean vectors, i.e., components of two-dimensional vectors over an orthonormal basis. Bold calligraphic symbols ($\mat{\amplification}, \mat{\jacobi}, \ldots$) refer to matrices, and cursive letters~($\geo{P}, \geo{F}, \geo{L},\ldots$) to time-like or null geodesics.

\section{Preliminary discussion: reference space-time and lenses}
\label{sec:preliminary_discussion}

\subsection{Rays and beams of light}

Let a source of light and an observer be placed in an arbitrary space-time. We call \emph{light beam} the set of light rays that connect the source to the observer; a beam may have several components if the source is multiply imaged. We assume that the wavelength of light is much smaller than any relevant length scale of the problem (eikonal approximation) so that light rays are null geodesics.

\subsection{Smooth and rough fields}

\begin{figure}[t]
    \centering
\begingroup%
  \makeatletter%
  \providecommand\color[2][]{%
    \errmessage{(Inkscape) Color is used for the text in Inkscape, but the package 'color.sty' is not loaded}%
    \renewcommand\color[2][]{}%
  }%
  \providecommand\transparent[1]{%
    \errmessage{(Inkscape) Transparency is used (non-zero) for the text in Inkscape, but the package 'transparent.sty' is not loaded}%
    \renewcommand\transparent[1]{}%
  }%
  \providecommand\rotatebox[2]{#2}%
  \newcommand*\fsize{\dimexpr\f@size pt\relax}%
  \newcommand*\lineheight[1]{\fontsize{\fsize}{#1\fsize}\selectfont}%
  \ifx\svgwidth\undefined%
    \setlength{\unitlength}{305.83825873bp}%
    \ifx\svgscale\undefined%
      \relax%
    \else%
      \setlength{\unitlength}{\unitlength * \real{\svgscale}}%
    \fi%
  \else%
    \setlength{\unitlength}{\svgwidth}%
  \fi%
  \global\let\svgwidth\undefined%
  \global\let\svgscale\undefined%
  \makeatother%
  \begin{picture}(1,0.43422148)%
    \lineheight{1}%
    \setlength\tabcolsep{0pt}%
    \put(0,0){\includegraphics[width=\unitlength,page=1]{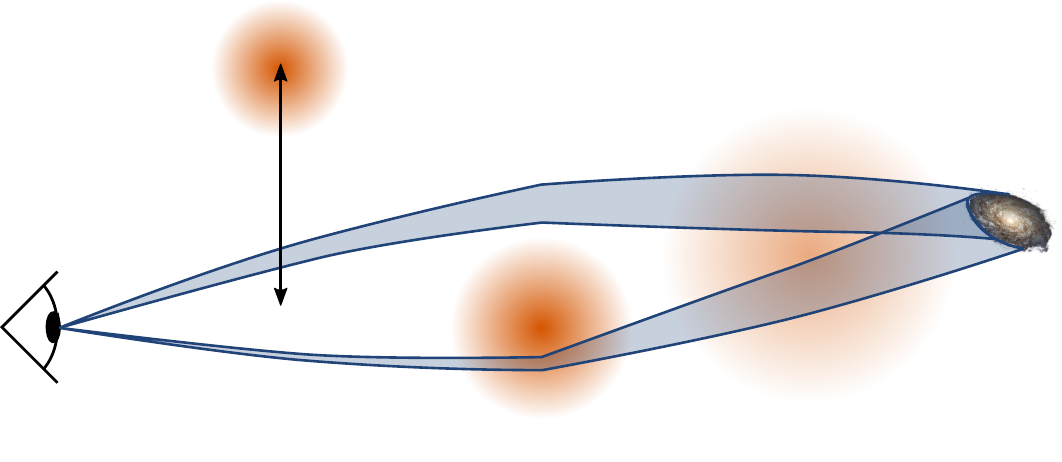}}%
    \put(0.13311841,0.24944043){\color[rgb]{0,0,0}\makebox(0,0)[lt]{\lineheight{1.25}\smash{\begin{tabular}[t]{l}$r\gg r\e{E}$\end{tabular}}}}%
    \put(0.42776228,0.36989894){\color[rgb]{0,0,0}\makebox(0,0)[lt]{\lineheight{1.25}\smash{\begin{tabular}[t]{l}smooth fields\end{tabular}}}}%
    \put(0.40781346,0.00810401){\makebox(0,0)[lt]{\lineheight{1.25}\smash{\begin{tabular}[t]{l}rough field\end{tabular}}}}%
    \put(0,0){\includegraphics[width=\unitlength,page=2]{smooth_vs_rough.pdf}}%
    \put(0.20575561,0.36247122){\color[rgb]{0,0,0}\makebox(0,0)[lt]{\lineheight{1.25}\smash{\begin{tabular}[t]{l}$r\e{E}$\end{tabular}}}}%
    \put(0,0){\includegraphics[width=\unitlength,page=3]{smooth_vs_rough.pdf}}%
    \put(0.58300831,0.16857613){\color[rgb]{0,0,0}\makebox(0,0)[lt]{\lineheight{1.25}\smash{\begin{tabular}[t]{l}$r\e{E}$\end{tabular}}}}%
    \put(0,0){\includegraphics[width=\unitlength,page=4]{smooth_vs_rough.pdf}}%
  \end{picture}%
\endgroup%

    \caption{Illustrating the dichotomy between smooth and rough gravity fields.}
    \label{fig:smooth_vs_rough}
\end{figure}

As a beam of light propagates from the source to the observer, it may be deflected, distorted, and somehow split, by the gravitational field that it experiences. The formalism proposed in this article relies on the broad distinction between two categories of gravitational fields: \emph{smooth fields} on the one hand, and \emph{rough fields} on the other hand. In smooth-field regions, the light beam is continuously (and possibly strongly) distorted, but its integrity is preserved. In rough-field regions, the light beam experiences sudden deflections, which may give rise to multiple images. Thus, rough-field regions correspond to what is usually referred to as \emph{lenses}, while smooth-field regions constitute a \emph{reference space-time}, where light propagates from one lens to the next one. In the standard multi-plane lensing framework, the reference space-time is either Minkowski or FLRW. We shall not make this assumption here.\footnote{Rigorously speaking, the set of all smooth-field regions traversed by a light beam does not constitute a physically well-defined space-time. In particular, it does not necessarily satisfy Einstein's equation, whose right-hand side should also include the matter clumps of the rough-field regions.} These considerations are illustrated in \cref{fig:smooth_vs_rough}.

Specifically, we shall say that a gravitational field is smooth if the light beam can be considered \emph{infinitesimal} within that field. In other words, a field is smooth if the corresponding space-time curvature is slowly varying and homogeneous within the light beam's cross section~\citep{Fleury:2017owg, Fleury:2018cro, Fleury:2018odh}. Otherwise, we shall say that the field is rough. Let us illustrate this terminology with the example of a point mass~$M$. The curvature that it produces at a distance $r$ reads $R^{\mu\nu\rho\sigma}R_{\mu\nu\rho\sigma}=12(2GM/r^3)^2$; hence the typical scale over which it changes appreciably is $r$. As a light beam with cross-sectional area~$A$ passes next to the mass, it experiences a smooth field if $A\ll r^2$, and a rough field otherwise. The picture gets slightly more complicated if we recall that the beam's cross-sectional area~$A(r)$ to depend on $r$, due to light focusing. Denoting $A_0$ the unlensed area of the beam, its lensed counterpart reads $A(r)=\mu(r) A_0$, where $\mu(r)\define (1-r\e{E}^4/r^4)^{-1}$ is the point-lens magnification, and $r\e{E}$ its Einstein radius. The smooth-field condition then becomes $A_0\ll r^2(1-r\e{E}^4/r^4)$.

Another example is light propagating through a diffuse distribution of matter, such as a gas cloud or a dark-matter halo. In that situation, space-time curvature is effectively dominated by its Ricci component~\citep{Fleury:2017owg}, which is mostly controlled by the matter density field $\rho$. Consequently, the corresponding gravitational field is smooth if $\rho$ is almost homogeneous on the scale of the beam's cross-section, and rough otherwise.

Finally, we note that gravitational fields are not always directly generated by matter distributions in a Newtonian-like manner. The most immediate example is gravitational waves, which are nothing but propagating curvature. In our terminology, a gravitational wave is a smooth field if its wavelength is much larger than the beam's diameter, and rough otherwise. Perhaps even more relevant to lensing are the infinite-wavelength gravitational waves that characterise anisotropic cosmologies. For instance, the anisotropic expansion of Bianchi~I models produces a homogeneous Weyl curvature, i.e. a smooth tidal field, which continuously shears and rotates light beams as they propagate~\citep{Fleury:2014rea}.

\subsection{Embedding lenses in the reference space-time}

In the remainder of this article, we aim to concretely implement the distinction between smooth and rough fields into a generalised multi-plane lensing framework. But before doing so, let us briefly explain how one may technically treat the embedding of a lens within an arbitrary reference space-time.

\begin{figure}[t]
    \centering
\begingroup%
  \makeatletter%
  \providecommand\color[2][]{%
    \errmessage{(Inkscape) Color is used for the text in Inkscape, but the package 'color.sty' is not loaded}%
    \renewcommand\color[2][]{}%
  }%
  \providecommand\transparent[1]{%
    \errmessage{(Inkscape) Transparency is used (non-zero) for the text in Inkscape, but the package 'transparent.sty' is not loaded}%
    \renewcommand\transparent[1]{}%
  }%
  \providecommand\rotatebox[2]{#2}%
  \newcommand*\fsize{\dimexpr\f@size pt\relax}%
  \newcommand*\lineheight[1]{\fontsize{\fsize}{#1\fsize}\selectfont}%
  \ifx\svgwidth\undefined%
    \setlength{\unitlength}{433.37960092bp}%
    \ifx\svgscale\undefined%
      \relax%
    \else%
      \setlength{\unitlength}{\unitlength * \real{\svgscale}}%
    \fi%
  \else%
    \setlength{\unitlength}{\svgwidth}%
  \fi%
  \global\let\svgwidth\undefined%
  \global\let\svgscale\undefined%
  \makeatother%
  \begin{picture}(1,0.3210844)%
    \lineheight{1}%
    \setlength\tabcolsep{0pt}%
    \put(0,0){\includegraphics[width=\unitlength,page=1]{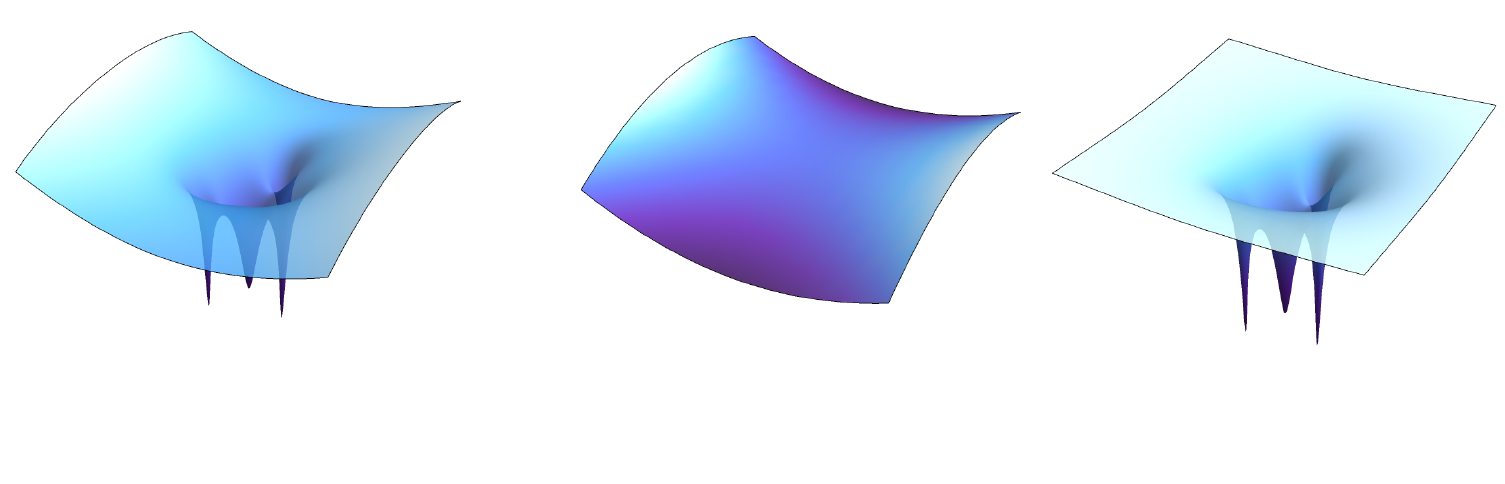}}%
    \put(0.03105236,0.04412093){\makebox(0,0)[lt]{\lineheight{1.25}\smash{\begin{tabular}[t]{l}complete space-time\end{tabular}}}}%
    \put(0.40970528,0.0456818){\makebox(0,0)[lt]{\lineheight{1.25}\smash{\begin{tabular}[t]{l}smooth reference\end{tabular}}}}%
    \put(0.82582504,0.04642026){\makebox(0,0)[lt]{\lineheight{1.25}\smash{\begin{tabular}[t]{l}lens\end{tabular}}}}%
    \put(0.31797002,0.04528268){\makebox(0,0)[lt]{\lineheight{1.25}\smash{\begin{tabular}[t]{l}$=$\end{tabular}}}}%
    \put(0.68767759,0.04636364){\makebox(0,0)[lt]{\lineheight{1.25}\smash{\begin{tabular}[t]{l}$+$\end{tabular}}}}%
    \put(0,0){\includegraphics[width=\unitlength,page=2]{embedding.pdf}}%
    \put(0.16557688,0.29169608){\makebox(0,0)[lt]{\lineheight{1.25}\smash{\begin{tabular}[t]{l}rough-field region\end{tabular}}}}%
    \put(0,0){\includegraphics[width=\unitlength,page=3]{embedding.pdf}}%
  \end{picture}%
\endgroup%

    \caption{Embedding a lens (rough-field region) in a smooth reference space-time.}
    \label{fig:embedding}
\end{figure}

Consider some rough-field region that is traversed by the light beam. In astrophysically relevant situations, the roughness of the field is due to a localised lumpy matter distribution (the lens), as depicted in \cref{fig:embedding}. Let $\geo{L}$ be the world-line of the lens's centre of mass. If we neglect the lens's self force, then $\geo{L}$ is a time-like geodesic of the space-time metric~$\bar{g}_{\mu\nu}$ generated by the rest of the universe, i.e., the reference space-time. Thus, we may introduce Fermi normal coordinates~$X^\alpha$ along $\geo{L}$, which materialise the \emph{rest frame of the lens}. With such coordinates, the reference metric is essentially Minkowskian across the rough-field region, $\bar{g}_{\alpha\beta}= \eta_{\alpha\beta}$, up to small corrections due to the local curvature of the reference space-time. These corrections are negligible in any astrophysically relevant situation, because the size of the region is on the order of the beam's cross section. Using the coordinates~$X^\alpha$, one may then compute the gravitational field generated by the lens as if it were isolated, modulo the aforementioned small curvature corrections.

In the remainder of this article, we shall assume that the lenses are non-compact matter distributions with non-relativistic velocity dispersion.  This notably excludes black-hole or neutron-star systems, and relativistic gases, but it provides an accurate description of any other astronomically relevant lens. In that context, we can treat rough gravitational fields as linear Newtonian perturbations, and hence apply the standard description of thin lenses~\citep{1992grle.book.....S}. Relaxing this assumption would affect the lens modelling and complicate the computation of time delays, but it would not change the essence of the framework that we propose in \cref{sec:one_lens,sec:N_lenses}.

\section{One lens in an arbitrary reference space-time}
\label{sec:one_lens}

In this section, we tackle the case of a light beam that only propagates in smooth-field regions, except one localised rough-field region modelled as a thin lens. This problem has been studied by several authors in order to evaluate the impact of cosmological perturbations on the properties of a strong lens. \citet{1987ApJ...316...52K} refers to it as a \emph{thick lens in the telescope approximation}; \citet{1992grle.book.....S, 1994A&A...287..349S} calls it \emph{generalised quadrupole lens}; \citet{1996ApJ...468...17B} writes about \emph{lensing with large-scale structure}; and \citet{Schneider:1997bq}, whose presentation is the closest to ours, simply calls it the \emph{cosmological lens}. The corresponding formalism has been recently applied to the weak lensing of Einstein rings by \citet{Birrer:2016xku}, with the perspective of novel synergies between weak-lensing and strong-lensing observations~\citep{Kuhn:2020wpy}.

The results derived in this section include, or are formally equivalent to, the aforementioned works'. However, we insert them in a broader context and extend their range of application --- a novel example is proposed in \cref{subsec:one_lens_Bianchi}. Furthermore, to our knowledge we propose in \cref{subsec:time_delay_one_lens} the first rigorous proof of the expression of the time delay for lensing within an arbitrary reference space-time.

\subsection{Geometric set-up}
\label{subsec:set-up}

Let us describe in detail the geometry of the problem. We shall start with a four-dimensional picture, which forms the rigorous basis of the subsequent three-dimensional picture, which in turn is more adapted to practical calculations. The four-dimensional discussion may be skipped by any reader who is not particularly interested in fully accurate definitions.

\begin{figure}[t]
\centering
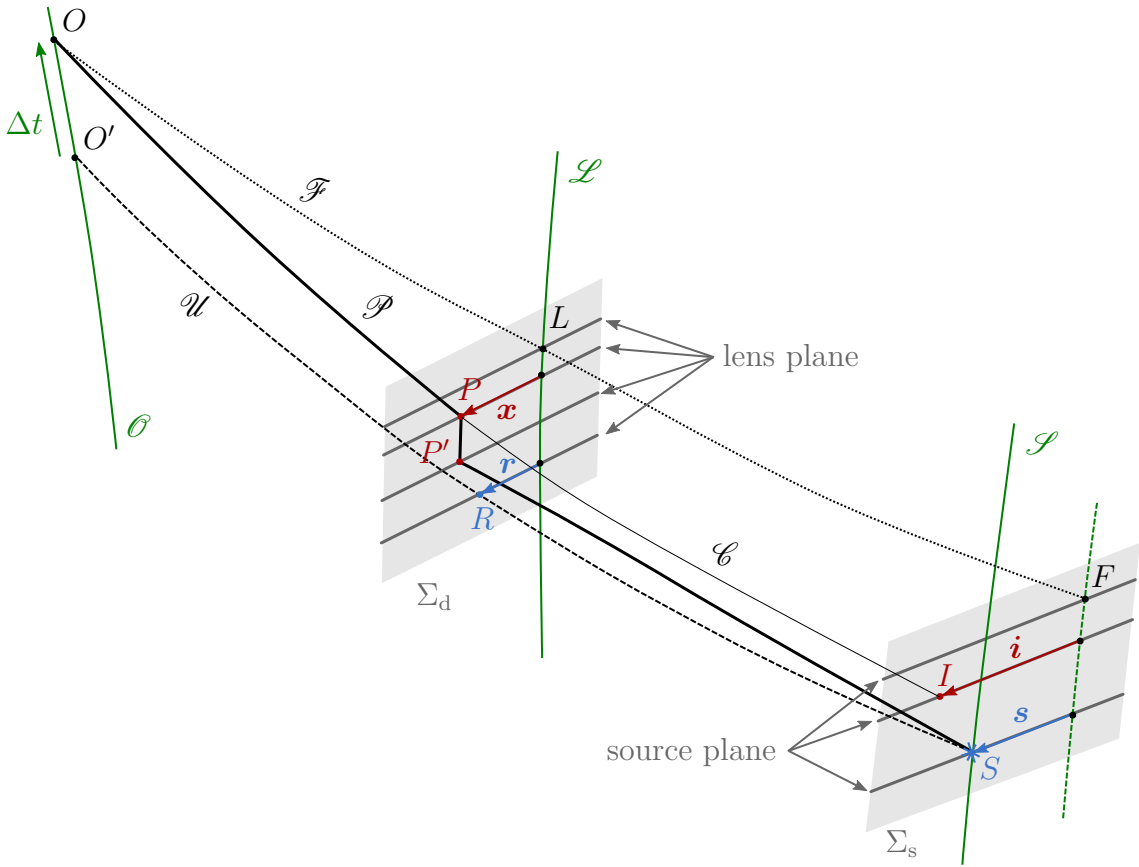
\caption{Space-time diagram for the lensing by a thin lens within an arbitrary smooth reference space-time. See the main text for precise definitions of the various objects. The green vertical lines~$\geo{O}, \geo{L}, \geo{S}$ are respectively the world-lines of the observer, lens and source. The solid thick line indicates the physical ray~$\geo{P}$, the dashed line is the unlensed ray~$\geo{U}$, the dotted line is the fiducial ray~$\geo{F}$, and the thin solid line is the continuation~$\geo{C}$ of the physical ray without deflection beyond the lens plane. The two grey shaded regions represent the lens and source world-sheets~$\Sigma\e{d}, \Sigma\e{s}$; their sections (thick grey lines) orthogonal to $\geo{L}$ or $\geo{S}$ indicate the lens and source planes at different times.}
\label{fig:one_lens_4d}
\end{figure}

\subsubsection{Four-dimensional picture (\cref{fig:one_lens_4d})}

Let $\geo{S}, \geo{L}, \geo{O}$ be the world-lines of the source, lens and observer respectively. Let $S\in\geo{S}$ and $O\in\geo{O}$ be the emission and observation events; we call \emph{physical ray}~$\geo{P}$ a null geodesic connecting $S$ to $O$. As we model the rough-field region as a thin lens, we will treat the physical ray as a set of two geodesics of the reference space-time, which undergoes a sudden deflection and a pause in the vicinity of the lens. The \emph{unlensed ray}~$\geo{U}$ is the null geodesic of the reference space-time that connects $S$ to $\geo{O}$. The intersection event $O'$ is earlier than $O$; their separation is the lensing time delay~$\Delta t$.

We introduce a \emph{fiducial ray} $\geo{F}$ defined as the geodesic of the reference space-time that connects $O$ to the lens's world-line~$\geo{L}$.\footnote{We make this choice for simplicity, but any reference-space-time geodesic that remains close to the physical ray could equally play the role of a fiducial ray.} We denote with $L\define\geo{F}\cap\geo{L}$ the intersection event. Importantly, we assume that the rays $\geo{P}, \geo{U}, \geo{F}$ are all very close to each other, so that any of their respective separations can be treated as infinitesimal in the reference space-time.

At $L$ we define the \emph{lens plane} as the two-dimensional space that is orthogonal to both $\geo{L}$ and $\geo{F}$. In other words, the lens plane is strictly spatial (made of simultaneous events) in the lens's rest frame, and it is orthogonal to the spatial direction of the fiducial ray in that frame. The lens plane is well defined in the immediate vicinity of $L$, i.e., up to distances that are much smaller than the curvature radii of the reference space-time. So far we have defined the lens plane at the time of the event $L$; we may then generalise it to other times through parallel transport along $\geo{L}$. Physically speaking, it means that the lens plane is non rotating. The three-dimensional time-like space that is swept by the lens plane as time goes on will be referred to as the \emph{lens world-sheet}~$\Sigma\e{d}$.

As the physical ray passes near the lens, it is effectively slowed down and deflected by its gravitational potential. In the thin-lens model adopted here, the deflection and delay are instantaneous. In other words, everything happens as if the photon were pausing and suddenly changing direction as it intersects $\Sigma\e{d}$. This pause is of course an idealised modelling of an otherwise continuous process, but it is an integral part of the thin lens approximation. We call $P'$ and $P$ the beginning and end of the pause, respectively. Since the pause happens at a fixed position in the lens's frame, the separation between $P$ and $P'$ is parallel to $\geo{L}$. The duration of the pause represents the so-called potential time delay;\footnote{For the thin-lens approximation to be valid, the duration of the pause must be very short compared to the local evolution time scale of the reference space-time.} it generally differs from the final time delay~$\Delta t$ measured along~$\geo{O}$, not only because these intervals are expressed in different rest frames, but also because $\Delta t$ contains additional, geometrical contributions.

We call $\vect{x}$ the (common) position of $P$ and $P'$ in the lens plane, with respect to the origin set by $\geo{L}$.\footnote{Throughout the article, bold symbols like $\vect{x}$ refer to the set of \emph{components} of screen-space vectors over an orthonormal basis. This allows us to treat them as Euclidean vectors.} Similarly, we call $\vect{r}$ the position, in the lens plane, of the intersection $R\define\geo{U}\cap\Sigma\e{d}$ between the unlensed ray and the lens world-sheet.

Let us now describe what happens in the vicinity of the source. Just like we defined the lens world-sheet~$\Sigma\e{d}$ from $\geo{F}$ and $\geo{L}$, we define the \emph{source world-sheet}~$\Sigma\e{s}$ from $\geo{F}$ and $\geo{S}$.\footnote{A slight difference is that $\geo{F}$ does not intersect~$\geo{S}$ in general. The explicit definition of $\Sigma\e{s}$ is the three-dimensional space that locally contains $\geo{S}$ and that intersects $\geo{F}$ orthogonally to its spatial direction in the source's frame.} Let us call $F\define\geo{F}\cap\Sigma\e{s}$ the intersection of the fiducial ray and the source world-sheet. The line parallel to $\geo{S}$ and passing through $F$ will be taken as the origin of the source plane. We call $\vect{s}$ the position of $S$ in the source plane with respect to that fiducial origin.

Finally, let $\geo{C}$ be the continuation of the physical ray without deflection nor delay in the lens plane. We call $I\define\geo{C}\cap\Sigma\e{s}$ its intersection with the source world-sheet. This event may be though of as the \emph{image event}, i.e. the event that would be observed at $O$ in the same direction as the physical ray in the absence of the lens. We denote with $\vect{i}$ the position of $I$ in the source plane.

The above discussion shows that, as long as the various rays can be considered infinitesimally close to each other in the reference space-time, we can univocally define the notions of lens plane and source plane, and the various position vectors~$\vect{x}, \vect{r}, \vect{s}, \vect{i}$ in those planes. We shall now safely proceed with a three-dimensional representation of the problem, which will allow us to represent angles more easily.

\subsubsection{Three-dimensional picture} 
\begin{figure}[t]
\centering
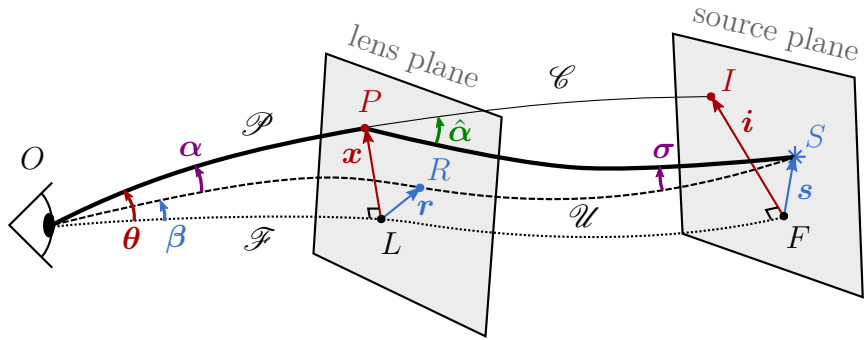
\caption{Three-dimensional representation of the lensing by a thin lens in an arbitrary smooth reference space-time. All the lines are portions of geodesics of the reference space-time. The unlensed ray~$\geo{U}$ (dashed line) connects the observer~$O$ to the source~$S$. Its angular separation with the arbitrary fiducial ray~$\geo{F}$ (dotted line) is denoted $\vect{\beta}$. The physical ray~$\geo{P}$ (thick solid line) is separated from the fiducial ray by $\vect{\theta}=\vect{
\beta}+\vect{\alpha}$; it is deflected by $\hat{\vect{\alpha}}$ at $P$ in the lens plane. The continued ray~$\geo{C}$ (light solid line) is the continuation of the physical ray if it were not deflected. The lens plane and source plane (grey) are orthogonal to the fiducial ray. The lens at $L$ is not represented in the figure.}
\label{fig:one_lens_3d}
\end{figure}

\Cref{fig:one_lens_3d}, which is a spatial representation of the space-time diagram of \cref{fig:one_lens_4d}, corresponds to the more traditional way of picturing gravitational lensing by a thin deflector. The fiducial ray~$\geo{F}$, which is a null geodesic of the reference space-time, was chosen to go through the lens's position for simplicity, but any other nearby ray would be eligible. The lens plane (resp. source plane) is perpendicular to the fiducial ray in the lens's (resp. source's) rest frame. The position of the source~$S$ with respect to the fiducial origin $F$ of the source plane is $\vect{s}$.

The unlensed ray~$\geo{U}$ is the geodesic of the reference space-time connecting $O$ to the $S$; we call $\vect{\beta}$ its angular separation with respect to the fiducial ray. Hence, $\vect{\beta}$ represents the direction in which the source would be observed in the reference space-time, without the lens. The unlensed ray intersects the lens plane at $R$, whose position with respect to $L$ is $\vect{r}$.

The physical ray~$\geo{P}$ is made of two portions of geodesics of the reference space-time, which intersect at $P$ in the lens plane. The position~$\vect{x}$ of $P$ with respect to $L$ is where the photon pauses and is deflected. We denote with $\hat{\vect{\alpha}}$ the \emph{deflection angle} of the physical ray at $P$. Importantly, $\hat{\vect{\alpha}}$ is defined in the rest frame of the lens; it is thus subject to aberration effects when evaluated in another frame.

We denote with $\vect{\theta}$ the separation between the physical ray and the fiducial ray at $O$, i.e. the observed direction of the image in the presence of the lens. The difference $\vect{\alpha}\define \vect{\theta}-\vect{\beta}$, not to be confused with $\hat{\vect{\alpha}}$, may be called \emph{displacement angle}. The analogue of $\vect{\alpha}$ in the source plane, i.e. the difference between the emission directions of the physical and unlensed rays, is denoted $\vect{\sigma}$.

Finally, the continued ray~$\geo{C}$ is the null geodesic of the reference space-time that coincides with the physical ray between $O$ and $P$. As such, it is not deflected at $P$ and intersects the source plane at a different point~$I$, called image position. It represents the position of a source that would be observed in the direction $\vect{\theta}$ in the absence of the lens. We call $\vect{i}$ the vector connecting $F$ to $I$ in the source plane.

All the angles~$\vect{\theta}, \vect{\beta}, \vect{\alpha}, \hat{\vect{\alpha}}, \vect{\sigma}$ are assumed to be very small.

\subsection{Lens equation for one lens}
\label{subsec:lens_equation_one_lens}

Now that the geometric set-up has been fully described, we are ready to derive the lens equation for one thin deflector embedded in the arbitrary reference space-time.

\subsubsection{Light propagation in the reference space-time} In the reference space-time, by definition, the rays~$\geo{F}, \geo{U}, \geo{P}, \geo{C}$ are considered infinitesimally close to each other. Thus, the relative behaviour of any two of these rays can be described by the Sachs framework \citep{1961RSPSA.264..309S}, which is based on geodesic deviation. In what follows, we shall use a number of results of this formalism without deriving them; we refer the interested reader to, e.g., \citet[][Chapt. 2]{Fleury:2015hgz} for further details.

Let $\lambda$ be a past-oriented affine parameter along the fiducial ray~$\geo{F}$, such that $\lambda=0$ at $O$. Just like we defined the lens plane and source plane, we may introduce a local screen space at any point~$\lambda$ of $\geo{F}$. Let us denote with $\vect{x}(\lambda)$ the screen-space separation between the physical and fiducial rays at $\lambda$. This vector field thus interpolates between $\vect{0}$ at $O$, $\vect{x}$ in the lens plane and $\vect{s}$ in the source plane.

In the reference space-time, $\vect{x}(\lambda)$ satisfies the differential equation
\begin{equation}
\label{eq:Sachs_equation}
\ddf[2]{\vect{x}}{\lambda} = \vect{\mathcal{R}}(\lambda) \, \vect{x}(\lambda) \ ,
\end{equation}
where $\vect{\mathcal{R}}$ is a particular screen-space projection of the Riemann curvature tensor of the reference space-time, called the optical tidal matrix. In fact, because \cref{eq:Sachs_equation} is linear, we immediately conclude that it would equally apply to the separation of \emph{any} two rays in the reference space-time.

\subsubsection{Jacobi matrices} \Cref{eq:Sachs_equation} is a linear second-order differential equation, so any of its solution is linearly related to its initial conditions. If this initial condition is considered at $O$ where $\vect{x}(0)=\vect{0}$, then there exists then there exists a $2\times 2$ matrix~$\mat{\jacobi}$, called \emph{Jacobi matrix}, such that $\vect{x}(\lambda)=\mat{\jacobi}(\lambda)\dot{\vect{x}}(0)$, where a dot denotes a derivative with respect to $\lambda$. The affine-parameter derivative~$\dot{\vect{x}}(0)$ of $\vect{x}$ at $O$ is related to the angle~$\vect{\theta}$ measured in the observer's rest frame as $\dot{\vect{x}}(0)=\omega_0\vect{\theta}$, with $\omega_0$ the cyclic frequency of light in that frame.\footnote{This relation is due to the most natural normalisation of the affine parameter. Let us denote with $u^\mu=\dd x^\mu/\dd\tau$ the four-velocity of an observer, with $\tau$ its proper time, and $k^\mu=\dd x^\mu/\dd\lambda$ the wave four-vector of the fiducial ray. It is customary identify the photon's cyclic frequency with $\omega=-u_\mu k^\mu=\dd\tau/\dd\lambda$. This normalisation condition implies that for a change $\dd\lambda$ the photon has travelled a proper distance $\dd\ell=\dd\tau=\omega\dd\lambda$ in the observer's frame.\label{footnote:distance_affine_parameter}}

More generally, if any two rays of the reference space-time emerge from, or converge to, a point (a)  with angular separation $\vect{\theta}\e{a}$, then their transverse separation~$\vect{x}\e{b}$ at another point (b) reads
\begin{equation}
\label{eq:Jacobi_definition}
\vect{x}\e{b}
= \mat{\jacobi}\e{a\langle b} \, \dot{\vect{x}}\e{a}
= \mat{\jacobi}\e{a\langle b} \, \omega\e{a}\vect{\theta}\e{a} \ ,
\end{equation}
where $\omega\e{a}$ is the fiducial photon's cyclic frequency as measured in the same rest frame where $\vect{\theta}\e{a}$ was defined.

The Jacobi matrix $\mat{\jacobi}\e{a\langle b}$ and the product $\omega\e{a} \vect{\theta}\e{a}$ are frame-independent. The presence of $\omega\e{a}$ is thus essential to account for aberration effects in $\vect{\theta}\e{a}$. We stress that a,b are not indices; they represent the positions where $\vect{x}, \vect{\theta}$ are evaluated.

The non-standard notation ``a$\langle$b'' in $\mat{\jacobi}\e{a\langle b}$ is designed to clearly indicate that the two rays merge at (a). If the roles of (a) and (b) were swapped, i.e. if we considered two rays merging at (b) instead of (a), then their separation at (a) would read $\vect{x}\e{a} = \mat{\jacobi}\e{a\rangle b} \, \omega\e{b} \vect{\theta}\e{b}$, with
\begin{equation}\label{eq:Etherington}
\mat{\jacobi}\e{a\rangle b} =  - \mat{\jacobi}\h{T}\e{a\langle b} \ ,
\end{equation}
where the T superscript indicates the matrix-transpose operation. \Cref{eq:Etherington} is known as Etherington's reciprocity law~\citep{1933PMag...15..761E}; see \citet[][\S~2.2.3]{Fleury:2015hgz} for a derivation using the same conventions as this article.

\subsubsection{Lens equation} From the definition of the Jacobi matrix, we can express the position of the image~$\vect{i}$ with respect to the source~$\vect{s}$ in two different ways,
\begin{equation}\label{eq:i-s_two_expressions}
\vect{i}-\vect{s}
= \omega\e{o}\mat{\jacobi}\e{o\langle s} \, \vect{\alpha}
= \omega\e{d}\mat{\jacobi}\e{d\langle s} \, \hat{\vect{\alpha}} \ ,
\end{equation}
where o, d, s respectively refer to the observer, deflector (or lens), and source positions. The deflection angle $\hat{\vect{\alpha}}(\vect{x})$ depends on the position $\vect{x}$ where the physical ray pierces the lens plane. For a thin lens made of non-relativistic and non-compact matter, $\hat{\vect{\alpha}}$ is dictated by the surface mass density~$\Sigma(\vect{x})$ in the lens plane~\citep{1992grle.book.....S},
\begin{equation}\label{eq:def_alpha}
\hat{\vect{\alpha}}(\vect{x})
= \int \dd^2\vect{y} \; 4G\Sigma(\vect{y}) \,
                        \frac{\vect{x}-\vect{y}}{|\vect{x}-\vect{y}|^2} \ ,
\end{equation}
where $|\ldots|$ denotes the Euclidean norm. The deflection angle can also be expressed as the gradient of $\hat{\psi}$, which is twice the projected gravitational potential of the lens,
\begin{equation}\label{eq:pot_lens}
\hat{\vect{\alpha}}(\vect{x})
= \ddf{\hat{\psi}}{\vect{x}} \ ,
\qquad
\hat{\psi}(\vect{x}) \define \int \dd^2\vect{y} \; 4G \Sigma(\vect{y}) \ln|\vect{x}-\vect{y}| \ .
\end{equation}
The latter indeed satisfies the projected Poisson equation $\Delta \hat{\psi}=8\pi G\Sigma(\vect{x})$, where $\Delta$ denotes the two-dimensional Laplacian and $G$ is Newton's constant.

Since $\vect{x}=\mat{\jacobi}\e{o\langle d} \omega\e{o} \vect{\theta}$, we conclude from \cref{eq:i-s_two_expressions} that the lens equation reads
\begin{empheq}[box=\fbox]{equation}
\label{eq:lens_equation_one_lens}
\vect{\beta}(\vect{\theta})
= \vect{\theta}
    - (1+z\e{d}) \mat{\jacobi}\e{o\langle s}^{-1}
				    \mat{\jacobi}\e{d\langle s}
					\hat{\vect{\alpha}}
					    (\mat{\jacobi}\e{o\langle d}\omega\e{o}\vect{\theta}) \ ,
\end{empheq}
with $z\e{d}$ the observed redshift of the lens, $1+z\e{d}=\omega\e{d}/\omega\e{o}$.

\subsubsection{Important remarks}

\Cref{eq:lens_equation_one_lens} is fully general, provided that light indeed encounters only one rough-field region, which can be modelled as a thin lens. No assumption is made on the reference space-time apart from the smoothness of its curvature. Hence, the deflector's redshift $z\e{d}$ must not be understood as a cosmological redshift in general.

Since $\hat{\vect{\alpha}}$ represents the deflection angle in the rest frame of the lens, it is by definition independent of the lens' motion. In that context, the redshift term $1+z\e{d}$ encodes aberration effects. For instance, if the lens recedes from the observer, then its redshift increases, and the net observed deflection $(1+z\e{d})\hat{\vect{\alpha}}$ increases as well.

Let us finally stress that $\vect{\beta}$ is fundamentally linked to the reference space-time. Indeed, $\vect{\beta}$ represents the direction in which one would observe the source without the lens, i.e., if light were only propagating in smooth-field regions. In particular, it does not represent the direction where the source would be seen in an empty Universe, because the smooth-field regions do affect light propagation. If one wishes to work with another reference space-time, which does not correspond to the geometry where light propagates from one lens to another (e.g. Minkowski or FLRW), then one must explicitly allow for the difference between $\vect{\beta}$ and the new notion of unlensed direction. This will be illustrated with two concrete examples in \cref{subsec:one_lens_perturbed_FLRW,subsec:one_lens_Bianchi}

\subsection{Amplification matrix for one lens}
\label{subsec:amplification_one_lens}

From the lens equation~\eqref{eq:lens_equation_one_lens}, we immediately derive the amplification matrix of the system, i.e. the Jacobian matrix of the lens mapping~$\vect{\theta}\mapsto\vect{\beta}(\vect{\theta})$,
\begin{equation}
\mat{\amplification}(\vect{\theta})
\define \ddf{\vect{\beta}}{\vect{\theta}}
= \mat{1} - \omega\e{d} \mat{\jacobi}\e{o\langle s}^{-1}
                        \mat{\jacobi}\e{d\langle s} \,
                        \hat{\mat{\hessian}}
                             (\mat{\jacobi}\e{o\langle d}\omega\e{o}\vect{\theta})\,
                        \mat{\jacobi}\e{o\langle d} \ ,
\end{equation}
where $\hat{\hessian}_{ij}\define \partial^2\hat{\psi}/\partial x^i \partial x^j$ is the Hessian matrix of the deflection potential. Note that, contrary to the latter, $\mat{\amplification}$ is generally not symmetric, due to the coupling between the lens and the reference space-time.

\subsection{Time delay for one lens}
\label{subsec:time_delay_one_lens}

The time delay between the images of strong-lensing systems is a key observable in astronomy and cosmology~\citep{2020MNRAS.498.1420W}. Its expression in the presence of cosmological perturbations can be found  in \citet{1992grle.book.....S, 1996ApJ...468...17B, Schneider:1997bq}, but without a direct proof. Here we propose a rigorous derivation of the time-delay formula, inspired from the wave-front method introduced by \citet[\S~5.3]{1992grle.book.....S}.

Let $\Delta t$ denote the time delay between the physical signal and the unlensed signal. In other terms, if a signal emitted by the source reached the observer at $t_0$ in the absence of the lens, then it would reach it at $t=t_0+\Delta t$ in the presence of the lens. Note that $\Delta t$ is not directly observable, because it involves two signals that propagate in different space-times. It is, however, a convenient theoretical intermediate.

The time delay is conveniently parameterised as $\Delta t(\vect{\theta}, \vect{\beta})$, because it generally depends on the source position $\vect{s}=\omega\e{o}\mat{\jacobi}\e{o\langle s}\vect{\beta}$, and on the point $\vect{x}=\omega\e{o}\mat{\jacobi}\e{o\langle d}\vect{\theta}$ where the physical ray pierces the lens plane. In terms of that parameterisation, the observable time delay between two images A and B of the same source reads $\Delta t\e{AB}(\vect{\beta})=t\e{A}(\vect{\beta})-t\e{B}(\vect{\beta})=\Delta t(\vect{\theta}\e{A},\vect{\beta})-\Delta t(\vect{\theta}\e{B},\vect{\beta})$.

\begin{figure}[t]
\centering
\begin{minipage}{0.3\textwidth}
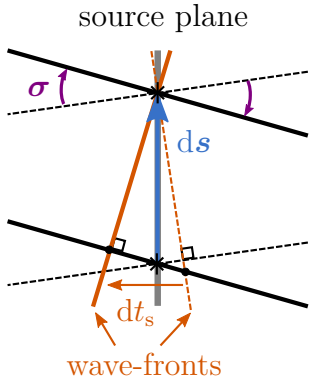
\end{minipage}
\hfill
\begin{minipage}{0.65\textwidth}
\caption{Two signals, emitted simultaneously with an angle $\vect{\sigma}$ from $\vect{s}+\dd\vect{s}$, are equivalent to two signals emitted with a relative delay $\dd t\e{s}=\vect{\sigma}\cdot\dd\vect{s}$ from $\vect{s}$. Here, everything happens as if the physical signal (solid) was emitted slightly before the unlensed signal (dashed), so that $\dd t\e{s}<0$ consistently with the opposite orientations of $\vect{\sigma}$ and $\dd\vect{s}$.}
\label{fig:time_delay_differential}
\end{minipage}
\end{figure}

\subsubsection{Differential time delay} The time delay between two signals emitted simultaneously in different directions depends on the source's position $\vect{s}$. Indeed, as depicted in \cref{fig:time_delay_differential}, if the source lies at $\vect{s}+\dd\vect{s}$, then everything happens as if the two signals were emitted from $\vect{s}$ but with a slight relative delay~$\dd t\e{s}=\vect{\sigma}\cdot\dd\vect{s}$. Thus, if two signals emitted from $\vect{s}$ are observed with a time delay $\Delta t$, then shifting the source by $\dd\vect{s}$ results in an additional delay
\begin{equation}
\label{eq:differential_time_delay}
\dd \Delta t = (1+z\e{s})\dd t\e{s} = (1+z\e{s})\vect{\sigma}(\vect{s})\cdot\dd\vect{s} \ ,
\end{equation}
where the redshift factor $1+z\e{s}=\dd t\e{o}/\dd t\e{s}$ allows for the fact that $\dd t\e{s}$ and $\dd\Delta t$ were defined in distinct frames.

\subsubsection{Time-delay formula} From \cref{eq:differential_time_delay}, we see that the expression of the time delay~$\Delta t$ may be obtained by a line integral of the vector field $\vect{\sigma}(\vect{s})$, between an arbitrary reference point and $\vect{s}$. In general, the line integral of a vector field depends on the path along which the integral is performed; except if the vector field is a gradient, which turns out to be the case. Namely, the emission angle $\vect{\sigma}$ between the lensed and unlensed rays depends on the source position $\vect{s}$ as
\begin{equation}
\label{eq:sigma_is_gradient}
(1+z\e{s})\vect{\sigma}(\vect{s}) = \ddf{T}{\vect{s}} \ ,
\end{equation}
where the scalar function $T$ reads
\begin{empheq}[box=\fbox]{align}
\label{eq:time_delay_one_lens}
T(\vect{\theta},\vect{\beta})
&\define \frac{1}{2} \, (\vect{\theta}-\vect{\beta})\cdot\mat{\tdmatrix}(\vect{\theta}-\vect{\beta})
	- (1+z\e{d})\hat{\psi}[\omega\e{o} \mat{\jacobi}\e{o\langle d} \vect{\theta})] \ ,
\\
\label{eq:time_delay_matrix}
\text{with}\quad
\mat{\tdmatrix} &\define \omega\e{o} \mat{\jacobi}\e{o\langle d}\h{T}\mat{\jacobi}\e{d\langle s}^{-1}\mat{\jacobi}\e{o\langle s} \ .
\end{empheq}
In \cref{eq:sigma_is_gradient}, the derivative $\dd T/\dd\vect{s}$ must be understood as a total derivative, in the sense that it accounts for the variation of \emph{both} natural variables $\vect{\theta}, \vect{\beta}$ of $T$ under a variation of $\vect{s}$. A detailed proof of \cref{eq:sigma_is_gradient} is provided in \cref{app:sigma_is_gradient}. Combining it with \cref{eq:differential_time_delay}, we immediately conclude that $\dd \Delta t = \dd T$, that is\footnote{Having noticed that $\vect{\sigma}(\vect{s})$ is a gradient makes our derivation of the time-delay formula simpler and more general than the one originally proposed in \citet{1992grle.book.....S}. In particular, we do not need to introduce a reference source whose contribution would be later set to zero on a caustic.}
\begin{empheq}[box=\fbox]{equation}
\Delta t(\vect{\theta}, \vect{\beta}) = T(\vect{\theta}, \vect{\beta}) + \cst \ .
\end{empheq}
Therefore, the time delay between two different images A, B of the same source reads $\Delta t\e{AB}(\vect{\beta})=T(\vect{\theta}\e{A},\vect{\beta})-T(\vect{\theta}\e{B},\vect{\beta})$. Note that any function of $\vect{\beta}$ could be added to the expression of $T(\vect{\theta},\vect{\beta})$ without changing the observable $\Delta t\e{AB}$.

The \emph{time-delay matrix}~$\mat{\tdmatrix}$ generalises the more common notion of time-delay distance to an arbitrary reference space-time. Indeed, if the latter is chosen as the FLRW space-time, then $\mat{\tdmatrix}=\bar{\tau}\,\mat{1}$, where $\bar{\tau}\define (1+z\e{d})\bar{D}\e{o\langle d}\bar{D}\e{o\langle s}/\bar{D}\e{d\langle s}$ is usually called the time-delay distance. Contrary to the Jacobi matrices that compose it, the time-delay matrix is symmetric,
\begin{equation}
\mat{\tdmatrix}\h{T} = \mat{\tdmatrix} \ .
\end{equation}
The time-delay matrix is related to, but different from, the telescope matrix introduced by \citet{1987ApJ...316...52K}, and then used by \citet{1992grle.book.....S, 1994A&A...287..349S, Schneider:1997bq} within the generalised quadrupole lens model. \citet{1987ApJ...316...52K, 1994A&A...287..349S, Schneider:1997bq} proved its symmetry with purely algebraic arguments. In \cref{app:symmetry_time_delay_matrix}, instead, we propose a geometric proof relying on Etherington's reciprocity law~\eqref{eq:Etherington}.

\subsubsection{Fermat's principle} Just like time-like and space-like geodesics can be defined from a stationary-time and stationary-length principle, null geodesics may be defined from Fermat's principle; see e.g. \citet{1992grle.book.....S}. In the present context, Fermat's principle states that the arrival time of a physical ray is stationary with respect to small variations of the position~$\vect{x}$ where it pierces the lens plane. In other words, all things being fixed (notably $\vect{s}, \vect{r}$), physical rays must satisfy $\partial T/\partial\vect{x}=\vect{0}$.

This property can be checked explicitly as follows. We first rewrite the first term of $T$ as $\vect{\alpha}\cdot\mat{\tdmatrix}\vect{\alpha}=(1+z\e{d})\hat{\vect{\alpha}}\cdot\hat{\mat{\tdmatrix}}\hat{\vect{\alpha}}$, where $\hat{\mat{\tdmatrix}}=\omega\e{d}\mat{\jacobi}\e{o\langle d}\mat{\jacobi}\e{o\langle s}^{-1}\mat{\jacobi}\e{d\langle s}$ is also a symmetric matrix. Then, using the identity $\hat{\mat{\tdmatrix}}\hat{\vect{\alpha}}=\vect{x}-\vect{r}$ we immediately find
\begin{equation}
\pd{T}{\vect{x}} = (1+z\e{d}) \pac{ \hat{\vect{\alpha}}(\vect{x}) - \ddf{\hat{\psi}}{\vect{x}} } ,
\end{equation}
so that physical light rays are indeed those whose deflection angle $\hat{\vect{\alpha}}$ is dictated by the physics of the lens plane.

\subsection{Example: one lens in a perturbed cosmological background}
\label{subsec:one_lens_perturbed_FLRW}

Suppose that the reference space-time can be treated as a weakly perturbed homogeneous-isotropic FLRW model. The associated Jacobi matrix takes the form
\begin{equation}
\label{eq:Jacobi_perturbed_FLRW}
\omega\e{a}\mat{\jacobi}\e{a\langle b}
= \bar{D}\e{a\langle b} \, \mat{\amplification}\e{a\langle b} \ .
\end{equation}
In \cref{eq:Jacobi_perturbed_FLRW}, $\bar{D}\e{a\langle b}=(1+z\e{b})^{-1}f_K(\chi\e{b}-\chi\e{a})$ denotes the angular diameter distance of (b) measured from (a) in the FLRW space-time, $\chi$ being the radial comoving distance, and $f_K(\chi)\define\sin(\sqrt{K}\chi)/\sqrt{K}$, with $K$ is the spatial-curvature parameter.

The second quantity,
\begin{equation}
\mat{\amplification}\e{a\langle b}
= \begin{bmatrix}
1-\kappa\e{a\langle b} - \Re(\gamma\e{a\langle b}) & - \Im(\gamma\e{a\langle b}) \\
- \Im(\gamma\e{a\langle b}) & 1-\kappa\e{a\langle b} + \Re(\gamma\e{a\langle b})
\end{bmatrix} ,
\end{equation}
is the amplification matrix due to cosmological perturbations about FLRW, i.e. caused by the large-scale matter inhomogeneities in the Universe. Its key components are the convergence~$\kappa\e{a\langle b}$ and complex shear~$\gamma\e{a\langle b}$; we have neglected the anti-symmetric part of $\mat{\amplification}\e{a\langle b}$, which represents the rotation of light beams with respect to parallel transport, because it is of the order of $\gamma^2$ if $\gamma\ll 1$ \citep[][\S~2.3.2]{Fleury:2015hgz}. At linear order in the matter density contrast~$\delta(\eta,\chi,\vect{x})\define(\rho-\bar{\rho})/\bar{\rho}$, the convergence and shear read~\citep{Fleury:2018cro}
\begin{align}
\label{eq:convergence}
\kappa\e{a\langle b}(\vect{\theta})
&= \frac{3}{2} \Omega\e{m0} H_0^2
	\int_{\chi\e{a}}^{\chi\e{b}} \dd\chi \; (1+z) \, \frac{f_K(\chi\e{b}-\chi)f_K(\chi-\chi\e{a})}{f_K(\chi\e{b}-\chi\e{a})}\,
	\delta[\eta_0-\chi, \chi, f_K(\chi)\vect{\theta}] \ , \\
\gamma\e{a\langle b}(\vect{\theta})
&= -\frac{3}{2} \Omega\e{m0} H_0^2
	\int_{\chi\e{a}}^{\chi\e{b}} \dd\chi \; (1+z) \, \frac{f_K(\chi\e{b}-\chi)f_K(\chi-\chi\e{a})}{f_K(\chi\e{b}-\chi\e{a})}
	\nonumber\\&\hspace{5cm} \times
	\int_{\mathbb{R}^2} \frac{\dd^2\vect{x}}{\pi x^2} \; \ex{2\ii\ph}\,
	    \delta[\eta_0-\chi, \chi, f_K(\chi)\vect{\theta}+\vect{x}] \ ,
\end{align}
where $\eta_0$ denotes today's conformal time, and $\ph$ is the polar angle of $\vect{x}=x(\cos\ph, \sin\ph)$.

\subsubsection{Lens equation}
\label{subsubsec:lens_equation_perturbed_FLRW}

In the lens equation~\eqref{eq:lens_equation_one_lens}, $\vect{\beta}$ stands for the direction in which a source would be observed without the lens, but still in the presence of the weak cosmological perturbations. In the lensing literature, however, it is understandably more common to write the lens equation in terms of the completely unlensed source position. That direction, which we shall denote $\bar{\vect{\beta}}=\mat{\amplification}\e{o\langle s}\vect{\beta}$, would be the observed position of the source without \emph{both} the strong lens and the cosmological perturbations, i.e. in an ideal FLRW Universe. In terms of $\bar{\vect{\beta}}$, \cref{eq:lens_equation_one_lens} reads
\begin{equation}
\label{eq:lens_equation_one_lens_cosmological_perturbations}
\bar{\vect{\beta}}(\vect{\theta})
= \mat{\amplification}\e{o\langle s}\vect{\theta}
- \frac{f_K(\chi\e{s}-\chi\e{d})} {f_K(\chi\e{s})}\,
	\mat{\amplification}\e{d\langle s} \,
	\hat{\vect{\alpha}}(\bar{D}\e{o\langle d}\mat{\amplification}\e{o\langle d}\vect{\theta}) \ .
\end{equation}
\Cref{eq:lens_equation_one_lens_cosmological_perturbations} had already been obtained in the literature under various forms; for instance, it is equivalent to eq.~(6.7) of \citet{1987ApJ...316...52K}, eq.~(14) of \citet{1996ApJ...468...17B}, eq.~(35) of \citet{McCully:2013fga}; it also coincides with eq.~(3.3) of \citet{Birrer:2016xku} in the critical-mass-sheet approximation defined therein.

\subsubsection{Equivalent lens}

In the absence of cosmological perturbations, i.e. for $\mat{\amplification}\e{a\langle b}=\vect{1}$, the lens equation would take the form
\begin{equation}\label{eq:one_lens_without_perturbation}
\bar{\vect{\beta}}=\vect{\theta}-\bar{\vect{\alpha}}(\vect{\theta}) \ ,
\end{equation}
with $\bar{\vect{\alpha}}(\vect{\theta})=[f_K(\chi\e{s}-\chi\e{d})/f_K(\chi\e{s})]\hat{\vect{\alpha}}(\bar{D}\e{\rm o\langle d}\vect{\theta})$. In the presence of perturbations, the lens equation~\eqref{eq:lens_equation_one_lens_cosmological_perturbations} is formally equivalent to \cref{eq:one_lens_without_perturbation}, if one replaces $\bar{\vect{\alpha}}(\vect{\theta})$ by
\begin{align}
\vect{\alpha}\e{eq}(\vect{\theta})
&= (\mat{1} - \mat{\amplification}\e{o\langle s})\vect{\theta}
    + \mat{\amplification}\e{d\langle s} \bar{\vect{\alpha}}
        (\mat{\amplification}\e{o\langle d}\vect{\theta})
\\
&\approx
    \pac{
        (\mat{1} - \mat{\amplification}\e{o\langle s})
        + \ddf{\bar{\vect{\alpha}}}{\vect{\theta}}
            (\mat{\amplification}\e{o\langle d}-\mat{1})
        } \vect{\theta}
    + \mat{\amplification}\e{d\langle s} \bar{\vect{\alpha}}(\vect{\theta}) \ ,
\end{align}
which may be called the equivalent lens. This shows that, in full generality, a lens model~$\bar{\vect{\alpha}}$ must be supplemented with 9 real parameters (3 $\kappa\e{a\langle b}$ and 3 complex $\gamma\e{a\langle b}$) to properly account for cosmological perturbations. Degeneracies between these parameters might occur if $\bar{\vect{\alpha}}$ enjoys symmetries.

In general, $\vect{\alpha}\e{eq}(\vect{\theta})$ cannot be written as a gradient, which means that it does not derive from a potential. An alternative approach \citep{Schneider:1997bq} which circumvents this issue consists in first applying a transformation $\vect{\beta}\mapsto\tilde{\vect{\beta}}\define \mat{\amplification}\e{o\langle d}\mat{\amplification}\e{d\langle s}^{-1}\vect{\beta}$ to the source plane. The resulting equivalent lens then derives from a potential.

\subsubsection{Amplification matrix}

Following the discussion of \cref{subsubsec:lens_equation_perturbed_FLRW}, we shall consider amplifications with respect to the homogeneous Universe, rather than the amplification due to the sole lens. The corresponding ``total'' amplification matrix is defined as $\mat{\amplification}\e{tot}=\dd\bar{\vect{\beta}}/\dd\vect{\theta}=\mat{\amplification}\e{o\langle s}\mat{\amplification}$, and reads
\begin{align}
\mat{\amplification}\e{tot}(\vect{\theta})
&= \mat{\amplification}\e{o\langle s}
	- \frac{f_K(\chi\e{s}-\chi\e{d})f_K(\chi\e{d})} {(1+z\e{d})f_K(\chi\e{s})} \,
	\mat{\amplification}\e{d\langle s} \,
	\hat{\mat{\hessian}}(\bar{D}\e{o\langle d}\mat{\amplification}\e{o\langle d}\vect{\theta}) \,
	\mat{\amplification}\e{o\langle d} \\
&= \mat{\amplification}\e{o\langle s}
	- \mat{\amplification}\e{d\langle s} \pac{\mat{1}-\overline{\mat{\amplification}}(\mat{\amplification}\e{o\langle d}\vect{\theta})} \mat{\amplification}\e{o\langle d} \ ,
\label{eq:total_amplification_one_lens_cosmological_perturbations}
\end{align}
where $\overline{\mat{\amplification}}\define\mat{1}- \dd\bar{\vect{\alpha}}/\dd\vect{\theta}$ is the amplification matrix that would characterise the lens in the absence of cosmological perturbations, i.e. in an ideal FLRW reference space-time. \Cref{eq:total_amplification_one_lens_cosmological_perturbations} confirms that line-of-sight perturbations do not only add to the effect of a lens, but they also modify the effect of the lens itself.

\subsubsection{Time delays}

For a perturbed FLRW reference space-time, the general expression~\eqref{eq:time_delay_one_lens} of the time delay becomes
\begin{align}
T(\vect{\theta},\bar{\vect{\beta}})
&= \frac{\bar{\tau}}{2} \, 
    (\vect{\theta}-\mat{\amplification}\e{o\langle s}^{-1}\bar{\vect{\beta}})\cdot
    \mat{\amplification}\e{o\langle d}
    \mat{\amplification}\e{d\langle s}^{-1}
    \mat{\amplification}\e{o\langle s}
    (\vect{\theta}-\mat{\amplification}\e{o\langle s}^{-1}\bar{\vect{\beta}})
- (1+z\e{d})\hat{\psi}(\bar{D}\e{o\langle d}\mat{\amplification}\e{o\langle d}\vect{\theta})
\\
&=
\bar{T}(\vect{\theta},\bar{\vect{\beta}})
+ \delta T(\vect{\theta},\bar{\vect{\beta}})
\end{align}
at first order in the cosmological perturbations, where, on the one hand
\begin{equation}
\bar{T}(\vect{\theta}, \bar{\vect{\beta}})
\define
\bar{\tau}
\pac{\frac{1}{2}
    |\vect{\theta}-\bar{\vect{\beta}}|^2
    - \bar{\psi}(\vect{\theta})
    }
\end{equation}
would be the time-delay function if the reference space-time were strictly homogeneous and isotropic, with $\bar{\tau}=(1+z\e{d}) \bar{D}\e{o\langle d}\bar{D}\e{o\langle s}/\bar{D}\e{d\langle s}$ the FLRW time-delay distance and $\bar{\psi}(\vect{\theta})= \bar{D}\e{d\langle s} \bar{D}\e{o\langle s}^{-1} \bar{D}\e{o\langle d}^{-1} \hat{\psi}(\bar{D}\e{o\langle d}\vect{\theta})$ the background lensing potential; on the other hand,\footnote{Note that we substituted the lens equation to obtain this expression of $\delta T$. A perhaps surprising consequence is that $\partial T/\partial\vect{\theta}\neq \vect{0}$ using that expression.}
\begin{equation}
\delta T(\vect{\theta},\bar{\vect{\beta}})
\define
\frac{1}{2} \, \bar{\tau}
(\vect{\theta}-\bar{\vect{\beta}}) \cdot
\pac{
    (\delta\mat{\amplification}\e{o\langle s}-
    \delta\mat{\amplification}\e{o\langle d})
    (\vect{\theta}+\bar{\vect{\beta}})
    -
    \delta\mat{\amplification}\e{d\langle s}
    (\vect{\theta}-\bar{\vect{\beta}})
    } ,
\end{equation}
where we denoted $\delta\mat{\amplification}\e{a\langle b}\define \mat{\amplification}\e{a\langle b}-\mat{1}$ for short, gathers all the corrections due to cosmological perturbations.

Taken at face value, the correction~$\delta T$ thereby induced is quite complex. However, for practical analyses of time-delay observations, these may be reduced to a \emph{single external convergence and shear}. First, since the source position~$\bar{\vect{\beta}}$ is unknown and hence a free parameter in such analyses, it does not make any difference whether one considers $\vect{\beta}=\mat{\amplification}\e{o\langle s}^{-1}\bar{\vect{\beta}}$ instead. Second, if the lens model $\bar{\psi}(\vect{\theta})$ is general enough,\footnote{In particular, an elliptic model may not suffice, since $\gamma\e{o\langle d}$ is generally not aligned with the intrinsic ellipticity of the lens.} then it may effectively account for the corrections due to $\mat{\amplification}\e{o\langle d}$ in the argument of $\hat{\psi}$. In that context, the time-delay model that must be used reads
\begin{equation}
T\e{mod}(\vect{\theta}, \vect{\beta})
=
\bar{\tau}
    \pac{\frac{1}{2}
        (\vect{\theta}-\vect{\beta})\cdot
        \mat{\amplification}\e{ext}
        (\vect{\theta}-\vect{\beta})
        - \psi\e{mod}(\vect{\theta})
        } ,
\end{equation}
with an ``external'' amplification matrix
\begin{equation}
\mat{\amplification}\e{ext}
\define
\mat{\amplification}\e{o\langle d}
\mat{\amplification}\e{d\langle s}^{-1}
\mat{\amplification}\e{o\langle s}
\approx
\mat{1} -
\begin{bmatrix}
\kappa\e{ext}+\Re(\gamma\e{ext}) & \Im(\gamma\e{ext}) \\
\Im(\gamma\e{ext})  & \kappa\e{ext}-\Re(\gamma\e{ext})
\end{bmatrix} ,
\end{equation}
featuring an external convergence
$
\kappa\e{ext} = \kappa\e{o\langle d} + \kappa\e{o\langle s} - \kappa\e{d\langle s}
$
and shear
$
\gamma\e{ext} = \gamma\e{o\langle d} + \gamma\e{o \langle s} - \gamma\e{d\langle s}
$.
While the external convergence is routinely implemented in current time-delay analyses~\citep{2020A&A...642A.194G}, we are not aware of any practical implementation of the external shear, although its relevance was suggested by \citet{McCully:2016yfe}.

\subsection{Example: lensing in an anisotropic Universe}
\label{subsec:one_lens_Bianchi}

As a second illustration, consider the case of a lens placed in a homogeneous but anisotropic Universe. If its homogeneity hyper-surfaces have no intrinsic curvature, then it may be described by the Bianchi~I geometry~\citep{1969CMaPh..12..108E}. In the Bianchi~I space-time, cosmic expansion is the same everywhere, but it is faster in some directions than in others. In comoving coordinates, its metric reads
\begin{equation}
\dd s^2 = a^2(\eta)
        \pac{ -\dd\eta^2
            + \ex{2\beta_x(\eta)}\dd x^2
            + \ex{2\beta_y(\eta)}\dd y^2
            + \ex{2\beta_z(\eta)}\dd z^2 } ,
\end{equation}
where $a(\eta)$ is the volume scale factor, and the three $\beta_i(\eta)$, which sum to zero, encode the anisotropy of expansion.

The propagation of light in Bianchi~I cosmologies has been thoroughly investigated in \citet{Fleury:2014rea, Fleury:2016htl}, thereby extending previous works on the same topic~\citep{saunders_observations_1969}. Let us restrict for simplicity to the weak-anisotropy limit ($\beta_i\ll 1$), where the optical properties of Bianchi~I reduce to those of a perturbed FLRW whose amplification matrix reads
\begin{align}
\mat{\amplification}\h{BI}\e{o\langle s}
= \mat{1} 
    + \mat{\mathcal{B}}(\eta\e{s})
    - \mat{\mathcal{B}}(\eta\e{o})
    - \frac{2}{\eta\e{o}-\eta\e{s}} \int_{\eta\e{s}}^{\eta\e{o}} \dd\eta
        \pac{\mat{\mathcal{B}}(\eta)+\frac{1}{2}\tr\mat{\mathcal{B}}(\eta)}
\ .
\end{align}
We used the matrix $\mathcal{B}_{AB}(\eta)\define \vect{s}_A\h{o} \cdot \mathrm{diag}[\beta_x(\eta), \beta_y(\eta), \beta_z(\eta)] \vect{s}_B\h{o}$ where $\vect{s}_1\h{o}, \vect{s}_2\h{o}$ denote the Sachs basis at the observer. Note that in $\mat{\amplification}\h{BI}\e{o\langle s}$ the source (s) and observer (o) are defined from their conformal time; this does not account for the change in redshift due to the anisotropic expansion, which reads $1+z=(1+\bar{z})[1+\tr\mat{\mathcal{B}}(\eta\e{s})-\tr\mat{\mathcal{B}}(\eta\e{o})]$.

All the results of \cref{subsec:one_lens_perturbed_FLRW} can be used to describe lensing in a weakly anisotropic Universe, if one uses the $\mat{\amplification}\h{BI}\e{o\langle s}$ as amplification matrix all along.

\section{$N$ lenses in an arbitrary reference space-time}
\label{sec:N_lenses}

\begin{figure}[t]
\centering
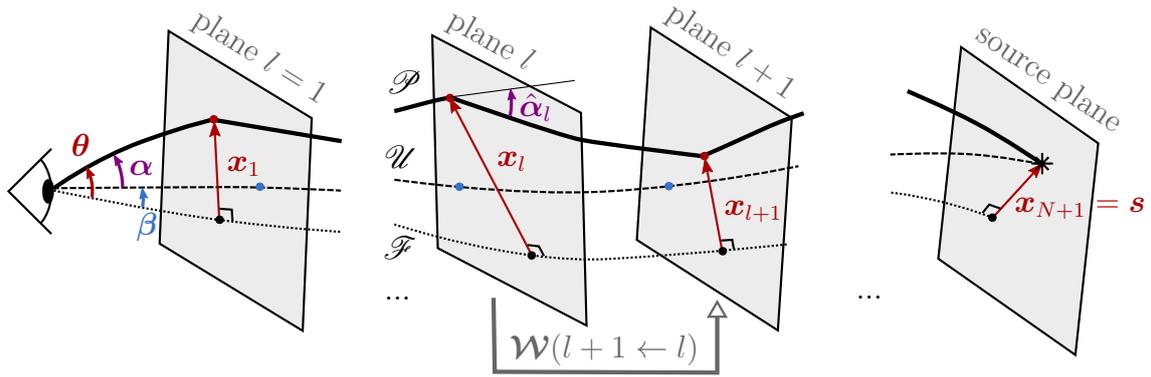
\caption{Same as \cref{fig:one_lens_3d}, but with $N$ lenses labelled by $l$. The transverse vector~$\vect{x}_l$ represents the position where the physical ray intersects the $l$th lens plane. The observer would correspond to $l=0$ ($\vect{x}_0=0$) and the source to $l=N+1$ ($\vect{x}_{N+1}=\vect{s}$).}
\label{fig:N_lenses}
\end{figure}

Let us now turn to the more involved case where light travels through an arbitrary number $N$ of rough-field regions. A three-dimensional representation of the set-up is depicted in \cref{fig:N_lenses}, where the various quantities are defined in the same way as in the one-lens case. To our knowledge, such a situation had never been considered in full generality, although the hybrid framework proposed by \citet{McCully:2013fga} may allow one to treat the most relevant cases in practice.

\subsection{Lens equation for $N$ lenses}

For $N$ lenses, the most direct derivation of the lens equation differs from the single-lens case in its structure. Let us introduce again a past-oriented affine parameter~$\lambda$ along $\geo{F}$, such that $\lambda=0$ at the observer and increases towards the source. We still denote with $\vect{x}(\lambda)$ the transverse separation between the physical and fiducial rays. Between two successive lens planes, $\vect{x}(\lambda)$ evolves smoothly according to the Sachs equation~\eqref{eq:Sachs_equation} of the reference space-time. When light reaches a lens plane, its sudden deflection implies a discontinuity of the derivative~$\dot{\vect{x}}\define\dd\vect{x}/\dd\lambda$.

\subsubsection{Deflection by a lens}

Let us denote $\Delta\dot{\vect{x}}_l\define\dot{\vect{x}}(\lambda_l^+)-\dot{\vect{x}}(\lambda_l^-)$ the discontinuity of $\dot{\vect{x}}$ in the $l$th plane. This quantity is not exactly the deflection angle $\hat{\vect{\alpha}}_l$, because the latter represents the discontinuity in the way $\vect{x}$ changes with \emph{proper distance}~$\ell$ in the lens' rest frame. As shown in \cref{footnote:distance_affine_parameter}, proper distance and affine parameter are related by $\dd\ell=\omega\dd\lambda$, so that
\begin{equation}
\Delta\dot{\vect{x}}_l
= \pa{\ddf{\ell}{\lambda}}_l \Delta\pa{\ddf{\vect{x}}{\ell}}_l
= - \omega_l \hat{\vect{\alpha}}_l \ ,
\end{equation}
where the minus sign comes from the conventional orientation of $\hat{\vect{\alpha}}$.

\subsubsection{Between two lenses}

Contrary to \cref{sec:one_lens}, the propagation from one lens plane to the next one does not necessarily involve merging rays; as such, it cannot be expressed in terms of Jacobi matrices only. However, we may still exploit the fact that $\vect{x}(\lambda)$ is the solution of a linear second-order differential equation~\eqref{eq:Sachs_equation}. In full generality, the state of that solution at $\lambda$ is encoded in the ``phase space'' vector
\begin{equation}
\vect{X}(\lambda) \define
\begin{bmatrix}
\vect{x}(\lambda)\\
\dot{\vect{x}}(\lambda)
\end{bmatrix} \ .
\end{equation}
Cauchy's theorem then implies the existence of a $4\times 4$ \emph{Wronski matrix}~$\mat{\wronski}$ such that for any $\lambda_1, \lambda_2$,
\begin{equation}
\label{eq:definition_Wronski}
\vect{X}(\lambda_2) = \mat{\wronski}(\lambda_2\leftarrow\lambda_1) \vect{X}(\lambda_1) \ .
\end{equation}
Although the Wronski matrix will not appear in the final lens equation, it is a very convenient tool for its derivation. Note that the top-right $2\times 2$ block of $\mat{\wronski}(\lambda_2\leftarrow\lambda_1)$ is actually the Jacobi matrix~$\mat{\jacobi}_{1\langle 2}$. Indeed, if two rays cross at $\lambda_1$, then $\vect{X}_1=(\vect{0},\dot{\vect{x}}_1)$, and $\vect{X}_2=(\mat{\jacobi}_{1\langle 2}\dot{\vect{x}}_1,\dot{\vect{x}}_2)$. Another important property of $\mat{\wronski}$, which trivially follows from its definition~\eqref{eq:definition_Wronski}, is the product law
\begin{equation}
\mat{\wronski}(\lambda_3\leftarrow\lambda_1)
= 
\mat{\wronski}(\lambda_3\leftarrow\lambda_2)
\mat{\wronski}(\lambda_2\leftarrow\lambda_1) \ .
\end{equation}

In what follows, we denote by $\mat{\wronski}(l+1\leftarrow l)$ the Wronski matrix that relates $\vect{X}_{l+1}^-\define\vect{X}(\lambda_{l+1}^-)$ before its deflection in the $(l+1)$th plane, to $\vect{X}_l^+\define\vect{X}(\lambda_l^+)$ after its deflection in the $l$th plane,
\begin{equation}
\label{eq:transfer_l_l+1}
\vect{X}_{l+1}^- = \mat{\wronski}(l+1\leftarrow l) \vect{X}_{l}^+ .
\end{equation}

\subsubsection{Recursion relation and lens equation} Since $\vect{x}(\lambda)$ is continuous at $\lambda_l$, the discontinuity of the phase-space vector $\vect{X}$ reads
\begin{equation}
\label{eq:discontinuity_X}
\Delta\vect{X}_l \define \vect{X}_l^+ - \vect{X}_l^-
=
\begin{bmatrix}
\Delta\vect{x}_l \\
\Delta\dot{\vect{x}_l}
\end{bmatrix}
=
\begin{bmatrix}
\vect{0}\\
-\omega_l\hat{\vect{\alpha}_l}
\end{bmatrix} \ .
\end{equation}
Denoting $\vect{X}_l\define\vect{X}_l^-$ (just before deflection at $\lambda_l$) for short, \cref{eq:transfer_l_l+1,eq:discontinuity_X} yield the recursion relation
$
\vect{X}_{l+1} = \mat{\wronski}(l+1\leftarrow l) (\vect{X}_l + \Delta\vect{X}_l)
$,
which is solved as
\begin{equation}
\vect{X}_l
= \mat{\wronski}(l\leftarrow\mathrm{o}) \vect{X}\e{o}
	+ \sum_{m=1}^{l-1} \mat{\wronski}(l\leftarrow m) \Delta\vect{X}_m \ .
\end{equation}
Finally, isolating the first two components, $\vect{x}_l$, of $\vect{X}_l$ in the above, noting that $\vect{X}_0=(\vect{0},\omega\e{o}\vect{\theta})$, and expressing the result in terms of $\vect{\beta}_l\define (\omega\e{o}\mat{\jacobi}_{\mathrm{o}\langle l})^{-1}\vect{x}_l$, we find
\begin{empheq}[box=\fbox]{equation}
\label{eq:recursion_beta_l}
\vect{\beta}_l = \vect{\theta}
- \sum_{m=1}^{l-1} (1+z_m)
	\mat{\jacobi}_{\mathrm{o}\langle l}^{-1} \mat{\jacobi}_{m\langle l}
	\hat{\vect{\alpha}}_m(\vect{x}_m)
\end{empheq}
for any $l$, which only involves Jacobi matrices. The angle $\vect{\beta}_l$ represents the direction in which a source at $\vect{x}_l$ in the $l$th plane would be observed in the absence of foreground lenses $m<l$. The case $l=N+1$, with $\vect{\beta}_{N+1}=\vect{\beta}$ yields the explicit lens equation
\begin{empheq}[box=]{equation}
\label{eq:lens_equation_N_lenses}
\vect{\beta}
= \vect{\theta}
	- \sum_{l=1}^N (1+z_l) \mat{\jacobi}\e{o\langle s}^{-1}\mat{\jacobi}_{l\langle\mathrm{s}} \, \hat{\vect{\alpha}}_l(\vect{x}_l) \ .
\end{empheq}
Note that the single-lens equation~\eqref{eq:lens_equation_one_lens} is recovered for $N=1$. \Cref{eq:recursion_beta_l} matches eq.~(10) of \citet{Schneider:2014vka}, derived for a perturbed FLRW reference space-time.

\subsection{Amplification matrix for $N$ lenses}

Just like the lens equation is a recursion relation, the amplification matrix for $N$ lenses takes a recursive form. From $\vect{\beta}_l$ we shall define the intermediate amplification matrix $\mat{\amplification}_l\define \dd\vect{\beta}_l/\dd\vect{\theta}$, which characterises the distortions of an infinitesimal source in the $l$th plane due to the foreground lenses~$m<l$. By differentiating \cref{eq:recursion_beta_l}, we find the recursion relation
\begin{equation}
\mat{\amplification}_l
= \mat{1}
	- \sum_{m=1}^{l-1} \omega_m
	    \mat{\jacobi}_{\mathrm{o}\langle l}^{-1}
	    \mat{\jacobi}_{m\langle l} \,
		\hat{\mat{\hessian}}_m(\vect{x}_m) \,
		\mat{\jacobi}_{\mathrm{o}\langle m}
		\mat{\amplification}_m \ ,
\end{equation}
where $\hat{H}^m_{ij}\define \partial\hat{\psi}_m/\partial x^i \partial x^j$, and with initial condition $\mat{\amplification}_{1}=\mat{1}$ since $\vect{\beta}_1=\vect{\theta}$. The complete amplification matrix, accounting for all the lenses, is $\mat{\amplification}\define\dd\vect{\beta}/\dd\vect{\theta}= \mat{\amplification}_{N+1}$.

\subsection{Time delay for $N$ lenses}

\begin{figure}[t]
\centering
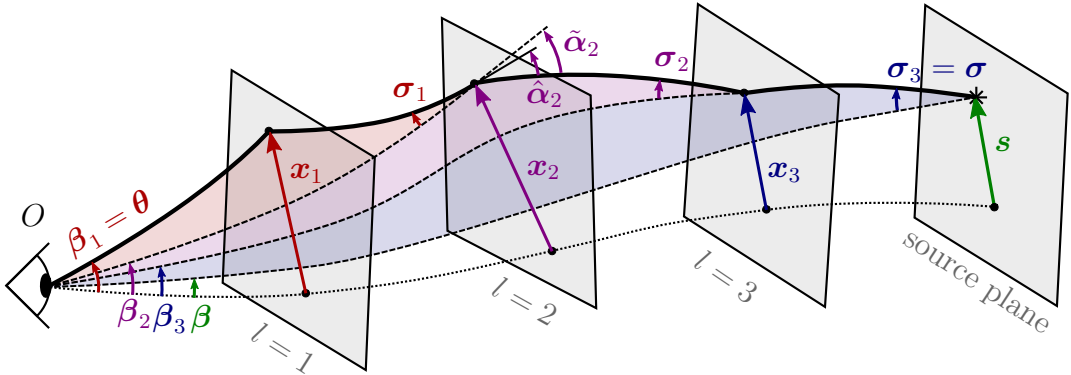
\caption{Same as \cref{fig:N_lenses} with $N=3$ lenses. Dashed lines represent rays of the reference space-time that connect the positions $\vect{x}_l$ of the physical ray to the observer.}
\label{fig:time_delay_N_lenses}
\end{figure}

\subsubsection{Expression of the time delay} The $N$-lens case can be intuitively deduced from the single-lens case as follows. First connect each point $\vect{x}_l$ to the observer with a fictitious ray of the reference space-time, as depicted in \cref{fig:time_delay_N_lenses} for $N=3$ lenses. We may then identify $N$ triangles formed by the points $O$, $\vect{x}_l$, $\vect{x}_{l+1}$. Let us apply the single-lens time-delay formula~\eqref{eq:time_delay_one_lens} in each of these triangles. Precisely, in the $l$th triangle, two signals emitted simultaneously at $\vect{x}_{l+1}$, the first one being undeflected (observed in the direction $\vect{\beta}_{l+1}$) and the other one deflected in the $l$th plane (observed in the direction $\vect{\beta}_{l}$), would be received with a delay
\begin{equation}
\label{eq:TdelayGen}
T_l(\vect{\beta}_l, \vect{\beta}_{l+1})
= \frac{1}{2}
    (\vect{\beta}_l-\vect{\beta}_{l+1})
    \cdot\mat{\tdmatrix}_{l(l+1)}
        (\vect{\beta}_l-\vect{\beta}_{l+1})
	- (1+z_l) \hat{\psi}_l(\vect{x}_l) \ ,
\end{equation}
up to a constant, with for any $l,m$ such that $0<l<m$,
\begin{equation}
\mat{\tdmatrix}_{lm}
\define 
    \omega\e{o} \mat{\jacobi}\h{T}_{\mathrm{o}\langle l}
                \mat{\jacobi}^{-1}_{l\langle m}
                \mat{\jacobi}_{\mathrm{o}\langle m} \ . 
\end{equation}
We note that the time-delay matrix satisfies the following unfolding relation;\footnote{\Cref{eq:unfolding_td_matrix} generalises eq.~(6.21) of \citet{2001stgl.book.....P}, which was also reported and exploited by \citet{McCully:2013fga}.};
for any three planes $l,m,n$ such that $0<l<m<n$,
\begin{equation}\label{eq:unfolding_td_matrix}
\mat{\tdmatrix}_{ln}^{-1} = \mat{\tdmatrix}_{lm}^{-1} + \mat{\tdmatrix}_{mn}^{-1} \ . 
\end{equation}
See \cref{app:unfold_td_matrix} for a proof. Although we shall not use it here, \cref{eq:unfolding_td_matrix} is very useful to derive the expression of time delays in the context of multi-plane lensing with a dominant lens~\citep{FLU21}.

The delay between the actual ray (observed in the direction $\vect{\theta}$) and the undeflected ray (observed in the direction $\vect{\beta}$) is then the sum of all these partial delays:
\begin{empheq}[box=\fbox]{align}
\label{eq:time_delay_N_lenses}
\Delta t &= T(\vect{\beta}_1,\ldots, \vect{\beta}_N)  + \cst \ ,
\\
T(\vect{\beta}_1,\ldots, \vect{\beta}_N) 
&\define \sum_{l=1}^N T_l(\vect{\beta}_l, \vect{\beta}_{l+1})  \ .
\end{empheq}

\subsubsection{Rigorous derivation} Although it yields the correct result, the above intuitive derivation of \cref{eq:time_delay_N_lenses} is actually incomplete. Indeed, we applied the single-lens time delay formula~\eqref{eq:time_delay_one_lens} to two non-physical rays. In particular, the ``deflection angle'' of the $l$th triangle, which is formed by the rays observed in the directions $\vect{\beta}_{l}, \vect{\beta}_{l+1}$, and which we may denote $\tilde{\vect{\alpha}}_l$, is \emph{not} the physical deflection angle $\hat{\vect{\alpha}}_l=\partial\hat{\psi}_l/\partial\vect{x}_l$ (see \cref{fig:time_delay_N_lenses}). This is not a mere detail, because the equality between the deflection angle and the gradient of the lensing potential was a key of the derivation of \cref{eq:time_delay_one_lens} proposed in \cref{app:sigma_is_gradient}. Therefore, nothing guarantees in principle that \cref{eq:time_delay_one_lens} can be applied to rays that do not exhibit the correct deflection angle.

Fortunately, despite this weakness in the way that \cref{eq:time_delay_N_lenses} was obtained, the formula itself turns out to be correct. Let us prove this point. First of all, we note that the differential expression~\eqref{eq:differential_time_delay}, i.e. $\dd\Delta t=(1+z_s)\vect{\sigma}\cdot\dd\vect{s}$, still applies here because it relies on strictly local arguments in the source plane. Therefore, if we can show that $(1+z\e{s})\vect{\sigma}=\dd T/\dd\vect{s}$, then it would imply $\dd\Delta t = \dd T$ just like in the single-lens case, and hence \cref{eq:time_delay_N_lenses} would follow.

Under a small variation of the source position $\vect{s}$, all the intermediate positions~$\vect{x}_l$ (and thus $\vect{\beta}_l$) change accordingly, so that each contribution $T_l$ of $T$ varies as
\begin{equation}
\ddf{T_l}{\vect{s}}
= \pa{\ddf{\vect{\beta}_l}{\vect{s}}}\h{T} \pd{T_l}{\vect{\beta}_l}
	+ \pa{\ddf{\vect{\beta}_{l+1}}{\vect{s}}}\h{T} \pd{T_l}{\vect{\beta}_{l+1}} \ .
\end{equation}
A similar calculation has already been performed in  \cref{app:sigma_is_gradient}, except that (i) what was denoted $\vect{s}$ there is now $\vect{x}_{l+1}$; and (ii) the deflection angle must be replaced by the geometrical angle $\tilde{\vect{\alpha}}_l$. In other words, \cref{eq:result_appendix_gradient} applied to the configuration of the $l$th triangle reads
\begin{equation}
\ddf{T_l}{\vect{x}_{l+1}}
= (1+z_l) \pa{ \ddf{\vect{x}_l}{\vect{x}_{l+1}} }\h{T}
				\pa{ \tilde{\vect{\alpha}}_l - \hat{\vect{\alpha}}_l }
	+ (1+z_{l+1}) \vect{\sigma}_l \ ,
\end{equation}
from which it is easy to get the derivative with respect to $\vect{s}$ by multiplying both sides with $(\dd\vect{x}_{l+1}/\dd\vect{s})\h{T}$; the variable with respect to which one differentiates is just a matter of parameterisation here. The last step consists in noticing the identity
\begin{equation}
\tilde{\vect{\alpha}}_l + \vect{\sigma}_{l-1} = \hat{\vect{\alpha}}_l  \ ,
\end{equation}
which clearly appears in \cref{fig:time_delay_N_lenses}, and from which we finally get
\begin{equation}
\ddf{T_l}{\vect{s}}
= (1+z_{l+1}) \pa{ \ddf{\vect{x}_{l+1}}{\vect{s}} }\h{T}  \vect{\sigma}_l
	- (1+z_l) \pa{ \ddf{\vect{x}_l}{\vect{s}} }\h{T} \vect{\sigma}_{l-1} \ .
\end{equation}

When summing the $\dd T_l/\dd\vect{s}$, all the terms cancel two by two, except the first one proportional to $\vect{\sigma}_0\define\vect{0}$, and the last one proportional to $\vect{\sigma}_N\define\vect{\sigma}$. Therefore,
\begin{equation}
\ddf{T}{\vect{s}} = \sum_{l=1}^N \ddf{T_l}{\vect{s}} = (1+z\e{s}) \vect{\sigma} \ ,
\end{equation}
which concludes our proof.

\subsubsection{Fermat's principle}

Like in the single-lens case, Fermat's principle states that a light ray is physical if and only if the time-delay function is stationary for this ray. Considering $T$ as a function of $\vect{x}_1, \ldots, \vect{x}_N$ instead of $\vect{\beta}_1, \ldots, \vect{\beta}_N$, one gets
\begin{equation}
\pd{T}{\vect{x}_l} = (1+z_l) \pac{ \hat{\vect{\alpha}}_l(\vect{x}_l) - \ddf{\hat{\psi}_l}{\vect{x}_l} }
\end{equation}
with similar calculations as in \cref{app:sigma_is_gradient}. Therefore, the function $T$ is stationary with respect to changes of $\vect{x}_1, \ldots, \vect{x}_N$ for and only for the physical ray.

\subsection{Example: lenses in an under-dense Universe}

As an illustration of the framework developed in this section, consider the situation where a fraction $f\in[0,1]$ of the matter in the Universe is homogeneously distributed, while the rest is under the form of lenses. Such a scenario is comparable to the Einstein-Straus Swiss-cheese model~\citep{1945RvMP...17..120E,1969ApJ...155...89K,Fleury:2013sna}, although in the latter the point-masses are surrounded by empty regions, while here we rather consider lenses placed within a homogeneous but under-dense cosmos. This under-dense Universe stands for our reference space-time, for which the Jacobi matrix is scalar,
\begin{equation}
\omega\e{a}\mat{\jacobi}\e{a\langle b} = D\e{a\langle b} \mat{1} \ ,
\end{equation}
where $D\e{a\langle b}$ is given by the Kantowski-Dyer-Roeder distance~\citep{1969ApJ...155...89K, 1973ApJ...180L..31D, 1973PhDT........17D, 1974ApJ...189..167D, Fleury:2014gha} with smoothness parameter $f$. For $z<2$, it may be approximated up to a few percent precision by the standard FLRW distance corrected by a negative convergence \citep{Kainulainen:2009dw}, $D\e{a\langle b}\approx \bar{D}\e{a\langle b}(1+\kappa\e{a\langle b})$, with
\begin{equation}
\kappa\e{a\langle b}
= \frac{3}{2} \Omega\e{m0} H_0^2 (f-1)
	\int_{\chi\e{a}}^{\chi\e{b}} \dd\chi \; (1+z) \, \frac{f_K(\chi\e{b}-\chi)f_K(\chi-\chi\e{a})}{f_K(\chi\e{b}-\chi\e{a})} \ , \\
\end{equation}
which was obtained from \cref{eq:convergence} with $\delta=f-1$.

In that setup, the lens recursion~\eqref{eq:recursion_beta_l} becomes
\begin{equation}
\vect{\beta}_l
= \vect{\theta}
- \sum_{m=1}^{l-1} \frac{D_{m\langle l}}{D_{\mathrm{o}\langle l}} \,
	\hat{\vect{\alpha}}_m(D_{\mathrm{o}\langle m}\vect{\beta}_m) \ ,
\end{equation}
which is identical to the original multi-plane recursion~\citep{1986ApJ...310..568B}, up to the expression of the distances. Such an approach efficiently meets the somehow heavier formalism developed by \citet[][\S~2.2.2]{McCully:2016yfe} to describe lenses separated by cosmic voids.

Time delays are also affected by the under-density of the reference space-time; the associated function is identical to the one of the standard multi-plane framework, 
\begin{align}
T(\vect{\beta}_1, \ldots \vect{\beta}_N)
&= \sum_{l=1}^N T_l(\vect{\beta}_l, \vect{\beta}_{l+1}) \ ,
\\
\text{with} \qquad
T_l(\vect{\beta}_l, \vect{\beta}_{l+1})
&= \frac{1}{2} \, \tau_{l(l+1)} \, |\vect{\beta}_{l+1}-\vect{\beta}_l|^2
    - (1+z_l)\hat{\psi}_l(D_{\mathrm{o}\langle l}\vect{\beta}_l) \ ,
\end{align}
except that the involved distances are changed, in particular
\begin{equation}
\tau_{l(l+1)}
= (1+z_l)
    \frac{D_{\mathrm{o}\langle l}D_{\mathrm{o}\langle l+1}}
        {D_{l\langle l+1}}
\approx \bar{\tau}_{l(l+1)}
        (1 + \kappa_{\mathrm{o}\langle l}
            + \kappa_{\mathrm{o}\langle l+1}
            - \kappa_{l\langle l+1}
        ) \ .
\end{equation}

\section{Conclusion}
\label{sec:conclusion}

In this article, we have proposed a comprehensive and efficient framework to model gravitational lensing by one or several deflectors placed in an arbitrary reference space-time. Our formalism relies on the dichotomy between \emph{smooth-field regions}, which form our \emph{reference space-time} where light beams can be considered infinitesimal, and \emph{rough-field regions} which can be described as isolated thin lenses. In that context, we have derived the lens equations, and the expressions of the amplification matrix and time delays. We illustrated our general results to: a single lens with cosmological perturbations; a single lens in an anisotropic Universe; and to $N$ lenses in an under-dense Universe.

In our derivations of the lens equation, amplification matrix and time delays, we have assumed that the lenses could be individually described by their Newtonian gravitational potential. This assumption is well motivated in most astrophysically relevant situations, but it does not apply to very hot systems, or in the vicinity of compact objects. Dealing with such relativistic lenses would require to change that specific part of the modelling, but it would not affect how lenses are embedded in the smooth reference space-time. 

Lensing by multiple deflectors had already been actively studied in the literature. The specific additions of the present work can be summarised as follows. In \cref{sec:one_lens}, we extended the description of a single lens with cosmological perturbations~\citep[e.g.][]{Schneider:1997bq} to a lens within any reference space-time. In particular, we proposed in \cref{subsec:time_delay_one_lens} the first rigorous derivation of the expression of time delays in that general context. \Cref{sec:N_lenses} further extended the results of \cref{sec:one_lens} to an arbitrary number of lenses, thereby generalising the standard multi-plane lensing formalism, which was hitherto limited to the Minkowski or FLRW reference space-times.

Our work establishes a firm basis for the description of gravitational lensing by several deflectors; in particular, it shall be applied to the accurate treatment of line-of-sight effects in strong gravitational lensing beyond the tidal approximation \citep[][in prep.]{FLU21}.

\section*{Acknowledgements}

PF thanks Sherry Suyu for kindly drawing his attention to the high-quality works of McCully et al., during a workshop in Benasque in 2019. We thank the anonymous referees of Classical and Quantum Gravity for their relevant comments, which improved the quality of our manuscript. We also thank Théo Duboscq and Daniel Johnson whose very careful reading revealed a few residual typos. PF received the support of a fellowship from ``la Caixa'' Foundation (ID 100010434). The fellowship code is LCF/BQ/PI19/11690018.

\appendix

\section{Derivation of \cref{eq:sigma_is_gradient}}
\label{app:sigma_is_gradient}

\begin{figure}[t]
\centering
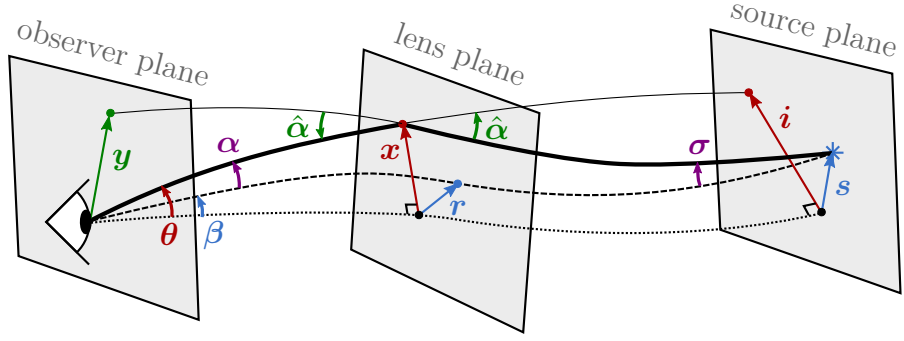
\caption{Same as \cref{fig:one_lens_3d}, except that we added the vector $\vect{y}$, which is obtained by continuing the physical ray without deflection from the source plane to the observer plane.}
\label{fig:one_lens_appendix}
\end{figure}

The goal of this appendix is to explicitly prove that the angle $\vect{\sigma}$ defined in \cref{subsec:time_delay_one_lens} can be written as a gradient,
\begin{align}
(1+z\e{s})\vect{\sigma}
&= \ddf{T}{\vect{s}} \ , \\
\text{where} \qquad
T &= \frac{1}{2}\, \vect{\alpha}\cdot\mat{\tdmatrix}\vect{\alpha}
			- (1+z\e{d})\hat{\psi}(\vect{x}) \ , \\
\text{and} \qquad
\mat{\tdmatrix}
&\define \omega\e{o} \mat{\jacobi}\e{o\langle d}\h{T}\mat{\jacobi}\e{d\langle s}^{-1}\mat{\jacobi}\e{o\langle s}
= \mat{\tdmatrix}\h{T} \ .
\end{align}
The set-up is reminded in \cref{fig:one_lens_appendix}, where we added the useful quantity $\vect{y}=-\omega\e{d}\mat{\jacobi}\e{d\langle o}\hat{\vect{\alpha}}=\omega\e{s}\mat{\jacobi}\e{s\langle o}\vect{\sigma}$ in the observer plane.

The two natural variables of $T$ being $\vect{\theta}, \vect{\beta}$, we decompose the total derivative as
\begin{equation}
\label{eq:total_derivative_T_s}
\ddf{T}{\vect{s}} = \pa{\ddf{\vect{\theta}}{\vect{s}}}\h{T} \pd{T}{\vect{\theta}}
								+ \pa{\ddf{\vect{\beta}}{\vect{s}}}\h{T} \pd{T}{\vect{\beta}} \ .
\end{equation}
Let us first compute $\partial T/\partial\vect{\theta}$ and $\partial T/\partial\vect{\beta}$. Thanks to the symmetry of $\mat{\tdmatrix}$, derivatives of $\vect{\alpha}\cdot\mat{\tdmatrix}\vect{\alpha}$ can be considered as hitting only the first $\vect{\alpha}$, while cancelling the pre-factor $1/2$, so that
\begin{align}
\pd{T}{\vect{\theta}} &= \mat{\tdmatrix}\vect{\alpha} - (1+z\e{d}) \ddf{\hat{\psi}}{\vect{\theta}} \ ,\\
\pd{T}{\vect{\beta}} &= - \mat{\tdmatrix}\vect{\alpha} \ .
\label{eq:derivative_T_beta}
\end{align}

To express $\mat{\tdmatrix}\vect{\alpha}$, we note from \cref{fig:one_lens_appendix} that the image displacement~$\vect{i}-\vect{s}$ can be written as a function of either $\vect{\alpha}$ or $\hat{\vect{\alpha}}$ as
$
\vect{i}-\vect{s} = \omega\e{o}\mat{\jacobi}\e{o\langle s}\vect{\alpha} = \omega\e{d}\mat{\jacobi}\e{d\langle s}\hat{\vect{\alpha}}
$. 
This yields
$
\mat{\jacobi}\e{d\langle s}^{-1}\mat{\jacobi}\e{o\langle s} \vect{\alpha}
= (1+z\e{d}) \hat{\vect{\alpha}}
$,
and thus
\begin{equation}
\label{eq:expression_T_alpha}
\mat{\tdmatrix}\vect{\alpha}
= \omega\e{o} \mat{\jacobi}\e{o\langle d}\h{T}\mat{\jacobi}\e{d\langle s}^{-1}\mat{\jacobi}\e{o\langle s} \vect{\alpha}
= \omega\e{d} \mat{\jacobi}\e{o\langle d}\h{T} \hat{\vect{\alpha}} \ .
\end{equation}
As $\vect{x}=\omega\e{o}\mat{\jacobi}\e{o\langle d}\vect{\theta}$, we get $\omega\e{o}\mat{\jacobi}\e{o\langle d}=\dd\vect{x}/\dd\vect{\theta}$, and hence the first term of \cref{eq:total_derivative_T_s} reads
\begin{equation}
\pa{\ddf{\vect{\theta}}{\vect{s}}}\h{T} \pd{T}{\vect{\theta}}
= (1+z\e{d}) \pa{\ddf{\vect{x}}{\vect{s}}}\h{T} \pa[4]{ \hat{\vect{\alpha}} - \ddf{\hat{\psi}}{\vect{x}} }
= \vect{0}
\end{equation}
for a physical ray. Therefore, the only non-zero contribution to $\dd T/\dd\vect{s}$ in \cref{eq:total_derivative_T_s} consists of the second term. From \cref{eq:derivative_T_beta,eq:expression_T_alpha} and $\vect{s}=\omega\e{o}\mat{\jacobi}\e{o\langle s}\vect{\beta}$, we have
\begin{align}
\pa{\ddf{\vect{\beta}}{\vect{s}}}\h{T} \pd{T}{\vect{\beta}}
&= - \pa{ \omega\e{o}\mat{\jacobi}\e{o\langle s}\h{T} }^{-1}
	\pa{ \omega\e{d} \mat{\jacobi}\e{o\langle d}\h{T} }
	\hat{\vect{\alpha}}
\\
&= - (1+z\e{s})
	\pa{ \omega\e{s}\mat{\jacobi}\e{o\rangle s} }^{-1}
	\pa{\omega\e{d} \mat{\jacobi}\e{o\rangle d}
	\hat{\vect{\alpha}} } \qquad \text{using \cref{eq:Etherington}}
\\
&= - (1+z\e{s})
	\pa{ \omega\e{s}\mat{\jacobi}\e{o\rangle s} }^{-1}
	(-\vect{y})
\\
&= (1+z\e{s}) \vect{\sigma} \ ,
\end{align}
where the last two steps may be followed with \cref{fig:one_lens_appendix}. We finally obtain
\begin{empheq}[box=\fbox]{equation}
\label{eq:result_appendix_gradient}
\ddf{T}{\vect{s}}
= (1+z\e{d}) \pa{\ddf{\vect{x}}{\vect{s}}}\h{T} \pa[4]{ \hat{\vect{\alpha}} - \ddf{\hat{\psi}}{\vect{x}} }
	+ (1+z\e{s}) \vect{\sigma}
= (1+z\e{s}) \vect{\sigma} \ ,
\end{empheq}
which concludes the proof. We chose to explicitly keep the term proportional to $\hat{\vect{\alpha}}-\dd\hat{\psi}/\dd\vect{x}$ in the result in order to apply it more easily to the $N$-lens case.

\section{Symmetry of the time-delay matrix}
\label{app:symmetry_time_delay_matrix}

Let us prove that the time-delay matrix
\begin{equation}
\mat{\tdmatrix}
\define
\omega\e{o}
\mat{\jacobi}\e{o\langle d}\h{T}
\mat{\jacobi}\e{d\langle s}^{-1}
\mat{\jacobi}\e{o\langle s}
=
- \omega\e{o}
\mat{\jacobi}\e{o\rangle d}
\mat{\jacobi}\e{d\langle s}^{-1}
\mat{\jacobi}\e{o\langle s}
\end{equation}
is symmetric, i.e., $\mat{\tdmatrix}=\mat{\tdmatrix}\h{T}$. For that purpose, we shall express the vector~$\vect{y}$ in the observer's plane (see \cref{fig:one_lens_appendix}) in two different ways. The proof will only rely on Etherington's reciprocity law, $\mat{\jacobi}\e{a\langle b}=-\mat{\jacobi}\e{a\rangle b}\h{T}$.

A first option consists in following the chain 
$
\vect{y}
\rightarrow\hat{\vect{\alpha}}
\rightarrow\vect{i}-\vect{s}
\rightarrow\vect{\alpha}
$, invoking the corresponding Jacobi matrices. The computation goes as follows,
\begin{align}
\vect{y}
&= -\omega\e{d}\mat{\jacobi}\e{o\rangle d}
    \hat{\vect{\alpha}}
\\
&= -\mat{\jacobi}\e{o\rangle d}
    \mat{\jacobi}\e{d\langle s}^{-1}
    (\vect{i}-\vect{s})
\\
&= -\mat{\jacobi}\e{o\rangle d}
    \mat{\jacobi}\e{d\langle s}^{-1}\,
    \omega\e{o}\mat{\jacobi}\e{o\langle s}
    \vect{\alpha}
\\
\label{eq:y_first_expression}
\vect{y}
&= \mat{\tdmatrix} \vect{\alpha}
\ .
\end{align}
The second option follows
$
\vect{y}
\rightarrow\vect{\sigma}
\rightarrow\vect{x}-\vect{r}
\rightarrow\vect{\alpha}
$
and reads
\begin{align}
\vect{y}
&= \omega\e{s}\mat{\jacobi}\e{o\rangle s}
    \vect{\sigma}
\\
&= \mat{\jacobi}\e{o\rangle s}
    \mat{\jacobi}\e{d\rangle s}^{-1}
    (\vect{x}-\vect{r})
\\
&= \mat{\jacobi}\e{o\rangle s}
    \mat{\jacobi}\e{d\rangle s}^{-1}\,
    \omega\e{o}\mat{\jacobi}\e{o\langle d}
    \vect{\alpha}
\\
&= -\omega\e{o}
    \mat{\jacobi}\e{o\langle s}\h{T}
    \pa{\mat{\jacobi}\e{d\langle s}^{-1}}\h{T}\,
    \mat{\jacobi}\e{o\rangle d}\h{T}
    \vect{\alpha}
\\
\label{eq:y_second_expression}
\vect{y}
&= \mat{\tdmatrix}\h{T} \vect{\alpha}
\ .
\end{align}
Comparing \cref{eq:y_first_expression,eq:y_second_expression}, which are valid for any $\vect{\alpha}$, concludes the proof.

\section{Unfolding relation~\eqref{eq:unfolding_td_matrix} for the time-delay matrix}
\label{app:unfold_td_matrix}

\begin{figure}[t]
\centering
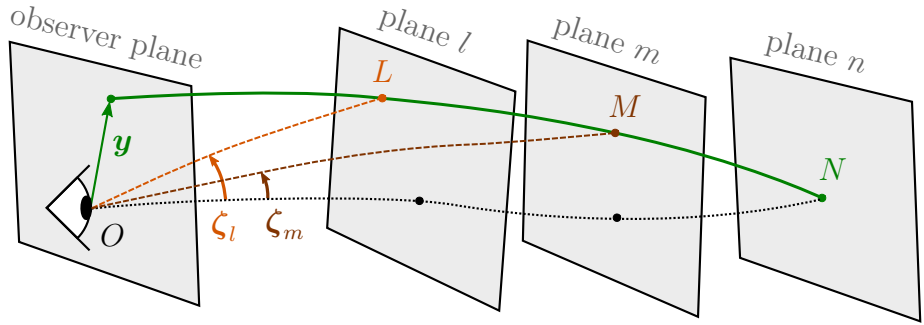
\caption{Geometrical meaning of the unfolding relation~\eqref{eq:unfolding_td_matrix} for the time-delay matrix. Every line is a geodesic of the reference space-time.}
\label{fig:unfolding_td_matrix}
\end{figure}

Let us prove that, for any three lens planes $l,m,n$ such that $0<l<m<n$, the time-delay matrix satisfies the unfolding relation
\begin{equation}\label{eq:unfolding_td_matrix_appendix}
\mat{\tdmatrix}_{ln}^{-1} = \mat{\tdmatrix}_{lm}^{-1} + \mat{\tdmatrix}_{mn}^{-1} \ .
\end{equation}
This relation is illustrated in \cref{fig:unfolding_td_matrix}, which is constructed as follows: (i) consider an arbitrary position~$\vect{y}$ in the observer plane; (ii) connect that position to the origin~$N$ of the $n$th lens plane with a ray of the reference space-time, and call $L$, $M$ the intersection between that ray with the two intermediate planes $l,m$; (iii) connect $L,M$ to the observer with rays of the reference space-time, which make angles $\vect{\zeta}_l, \vect{\zeta}_m$ with respect to the fiducial ray.

Proceeding just like in \cref{app:symmetry_time_delay_matrix}, we may then relate $\vect{y}$ to the angles $\vect{\zeta}_l, \vect{\zeta}_m$ as
\begin{align}
\vect{y} &= \mat{\tdmatrix}_{ln} \vect{\zeta}_l
\qquad\text{in triangle $(OLN)$,}\\
\vect{y} &= \mat{\tdmatrix}_{mn} \vect{\zeta}_m
\qquad\text{in triangle $(OMN)$,}\\
\vect{y} &= \mat{\tdmatrix}_{lm} (\vect{\zeta}_l-\vect{\zeta}_m)
\qquad\text{in triangle $(OLM)$.}
\end{align}
It follows that
$(\mat{\tdmatrix}_{ln}^{-1}-\mat{\tdmatrix}_{lm}^{-1}-\mat{\tdmatrix}_{mn}^{-1})\vect{y}=\vect{0}$, and therefore \cref{eq:unfolding_td_matrix_appendix} because the above holds for any $\vect{y}$ in the observer plane.

\bibliographystyle{mybibstyle}
\bibliography{bibliography.bib}

\begin{thebibliography}{}

\bibitem[Akrami et~al.(2020)]{Akrami:2019bkn}
{\bfseries Planck} Collaboration, Akrami, Y. et~al., ``{Planck 2018 results.
  VII. Isotropy and Statistics of the CMB}'', {\em Astron. Astrophys.}
  {\bfseries 641} (2020) A7,
  \href{https://arxiv.org/abs/1906.02552}{{\ttfamily [arXiv:1906.02552]}}.

\bibitem[Asgari et~al.(2021)]{Asgari:2020wuj}
{\bfseries KiDS} Collaboration, Asgari, M. et~al., ``{KiDS-1000 Cosmology:
  Cosmic shear constraints and comparison between two point statistics}'', {\em
  Astron. Astrophys.} {\bfseries 645} (2021) A104,
  \href{https://arxiv.org/abs/2007.15633}{{\ttfamily [arXiv:2007.15633]}}.

\bibitem[{Bar-Kana}(1996)]{1996ApJ...468...17B}
{Bar-Kana}, R., ``{Effect of Large-Scale Structure on Multiply Imaged
  Sources}'', {\em \apj} {\bfseries 468} (1996) 17,
  \href{https://arxiv.org/abs/astro-ph/9511056}{{\ttfamily
  [arXiv:astro-ph/9511056]}}.

\bibitem[Birrer et~al.(2017)Birrer, Welschen, Amara, and
  Refregier]{Birrer:2016xku}
Birrer, S., Welschen, C., Amara, A., and Refregier, A., ``{Line-of-sight
  effects in strong lensing: Putting theory into practice}'', {\em JCAP}
  {\bfseries 1704} (2017), no.~04, 049,
 \href{https://arxiv.org/abs/1610.01599}{{\ttfamily [arXiv:1610.01599]}}.

\bibitem[{Blandford} and {Narayan}(1986)]{1986ApJ...310..568B}
{Blandford}, R. and {Narayan}, R., ``{Fermat's principle, caustics, and the
  classification of gravitational lens images}'', {\em \apj} {\bfseries 310}
  (1986) 568--582.

\bibitem[Blum et~al.(2020)Blum, Castorina, and Simonovi\'c]{Blum:2020mgu}
Blum, K., Castorina, E., and Simonovi\'c, M., ``{Could Quasar Lensing Time
  Delays Hint to a Core Component in Halos, Instead of $H_0$ Tension?}'', {\em
  Astrophys. J. Lett.} {\bfseries 892} (2020), no.~2, L27,
  \href{https://arxiv.org/abs/2001.07182}{{\ttfamily [arXiv:2001.07182]}}.

\bibitem[\c{C}a\u{g}an \c{S}eng\"{u}l et~al.(2020)\c{C}a\u{g}an \c{S}eng\"{u}l,
  Tsang, Diaz~Rivero, Dvorkin, Zhu, and Seljak]{Sengul:2020yya}
\c{C}a\u{g}an \c{S}eng\"{u}l, A., Tsang, A., Diaz~Rivero, A., Dvorkin, C., Zhu,
  H.-M., and Seljak, U., ``{Quantifying the line-of-sight halo contribution to
  the dark matter convergence power spectrum from strong gravitational
  lenses}'', {\em Phys. Rev. D} {\bfseries 102} (2020), no.~6, 063502,
  \href{https://arxiv.org/abs/2006.07383}{{\ttfamily [arXiv:2006.07383]}}.

\bibitem[{Despali} et~al.(2018){Despali}, {Vegetti}, {White}, {Giocoli}, and
  {van den Bosch}]{2018MNRAS.475.5424D}
{Despali}, G., {Vegetti}, S., {White}, S. D.~M., {Giocoli}, C., and {van den
  Bosch}, F.~C., ``{Modelling the line-of-sight contribution in substructure
  lensing}'', {\em \mnras} {\bfseries 475} (2018), no.~4, 5424--5442,
  \href{https://arxiv.org/abs/1710.05029}{{\ttfamily [arXiv:1710.05029]}}.

\bibitem[Diego et~al.(2018)]{Diego:2017drh}
Diego, J.~M. et~al., ``{Dark Matter under the Microscope: Constraining Compact
  Dark Matter with Caustic Crossing Events}'', {\em Astrophys. J.} {\bfseries
  857} (2018), no.~1, 25,  \href{https://arxiv.org/abs/1706.10281}{{\ttfamily
  [arXiv:1706.10281]}}.

\bibitem[{Dyer}(1973)]{1973PhDT........17D}
{Dyer}, C.~C., ``{Observational Aspects of Locally Inhomogeneous Cosmological
  Models.}'', PhD thesis, University of Toronto (Canada), 1973.

\bibitem[{Dyer} and {Roeder}(1973)]{1973ApJ...180L..31D}
{Dyer}, C.~C. and {Roeder}, R.~C., ``{Distance-Redshift Relations for Universes
  with Some Intergalactic Medium}'', {\em \apjl} {\bfseries 180} (1973) L31.

\bibitem[{Dyer} and {Roeder}(1974)]{1974ApJ...189..167D}
{Dyer}, C.~C. and {Roeder}, R.~C., ``{Observations in Locally Inhomogeneous
  Cosmological Models}'', {\em \apj} {\bfseries 189} (1974) 167--176.

\bibitem[{Einstein} and {Straus}(1945)]{1945RvMP...17..120E}
{Einstein}, A. and {Straus}, E.~G., ``{The Influence of the Expansion of Space
  on the Gravitation Fields Surrounding the Individual Stars}'', {\em Reviews
  of Modern Physics} {\bfseries 17} (1945), no.~2-3, 120--124.

\bibitem[{Ellis} and {MacCallum}(1969)]{1969CMaPh..12..108E}
{Ellis}, G.~F.~R. and {MacCallum}, M.~A.~H., ``{A class of homogeneous
  cosmological models}'', {\em Communications in Mathematical Physics}
  {\bfseries 12} (1969), no.~2, 108--141.

\bibitem[{Etherington}(1933)]{1933PMag...15..761E}
{Etherington}, I.~M.~H., ``{On the Definition of Distance in General
  Relativity.}'', {\em Philosophical Magazine} {\bfseries 15} (1933) 761.

\bibitem[Fleury(2014)]{Fleury:2014gha}
Fleury, P., ``{Swiss-cheese models and the Dyer-Roeder approximation}'', {\em
  JCAP} {\bfseries 06} (2014) 054,
  \href{https://arxiv.org/abs/1402.3123}{{\ttfamily [arXiv:1402.3123]}}.

\bibitem[Fleury(2015)]{Fleury:2015hgz}
Fleury, P., ``{Light propagation in inhomogeneous and anisotropic
  cosmologies}'', PhD thesis, Paris U., VI, IAP, 2015.
\newblock  \href{https://arxiv.org/abs/1511.03702}{{\ttfamily
  [arXiv:1511.03702]}}.

\bibitem[Fleury et~al.(2013)Fleury, Dupuy, and Uzan]{Fleury:2013sna}
Fleury, P., Dupuy, H., and Uzan, J.-P., ``{Interpretation of the Hubble diagram
  in a nonhomogeneous universe}'', {\em Phys. Rev. D} {\bfseries 87} (2013),
  no.~12, 123526,  \href{https://arxiv.org/abs/1302.5308}{{\ttfamily
  [arXiv:1302.5308]}}.

\bibitem[Fleury et~al.(2017)Fleury, Larena, and Uzan]{Fleury:2017owg}
Fleury, P., Larena, J., and Uzan, J.-P., ``{Weak gravitational lensing of
  finite beams}'', {\em Phys. Rev. Lett.} {\bfseries 119} (2017), no.~19,
  191101,  \href{https://arxiv.org/abs/1706.09383}{{\ttfamily
  [arXiv:1706.09383]}}.

\bibitem[Fleury et~al.(2019a)Fleury, Larena, and Uzan]{Fleury:2018cro}
Fleury, P., Larena, J., and Uzan, J.-P., ``{Cosmic convergence and shear with
  extended sources}'', {\em Phys. Rev. D} {\bfseries 99} (2019)a, no.~2,
  023525,  \href{https://arxiv.org/abs/1809.03919}{{\ttfamily
  [arXiv:1809.03919]}}.

\bibitem[Fleury et~al.(2019b)Fleury, Larena, and Uzan]{Fleury:2018odh}
Fleury, P., Larena, J., and Uzan, J.-P., ``{Weak lensing distortions beyond
  shear}'', {\em Phys. Rev. D} {\bfseries 99} (2019)b, no.~2, 023526,
  \href{https://arxiv.org/abs/1809.03924}{{\ttfamily [arXiv:1809.03924]}}.

\bibitem[Fleury et~al.(2021)Fleury, Larena, and Uzan]{FLU21}
Fleury, P., Larena, J., and Uzan, J.-P., ``{Line-of-sight effects in strong
  gravitational lensing}'', 2021, in preparation.

\bibitem[Fleury et~al.(2016)Fleury, Nugier, and Fanizza]{Fleury:2016htl}
Fleury, P., Nugier, F., and Fanizza, G., ``{Geodesic-light-cone coordinates and
  the Bianchi I spacetime}'', {\em JCAP} {\bfseries 06} (2016) 008,
  \href{https://arxiv.org/abs/1602.04461}{{\ttfamily [arXiv:1602.04461]}}.

\bibitem[Fleury et~al.(2015)Fleury, Pitrou, and Uzan]{Fleury:2014rea}
Fleury, P., Pitrou, C., and Uzan, J.-P., ``{Light propagation in a homogeneous
  and anisotropic universe}'', {\em Phys. Rev. D} {\bfseries 91} (2015), no.~4,
  043511,  \href{https://arxiv.org/abs/1410.8473}{{\ttfamily
  [arXiv:1410.8473]}}.

\bibitem[Fosalba and Gaztañaga(2020)]{Fosalba:2020gls}
Fosalba, P. and Gaztañaga, E., ``{Explaining Cosmological Anisotropy: Evidence
  for Causal Horizons from CMB data}'',
  \href{https://arxiv.org/abs/2011.00910}{{\ttfamily [arXiv:2011.00910]}}.

\bibitem[Gatti et~al.(2020)]{Gatti:2019clj}
{\bfseries DES} Collaboration, Gatti, M. et~al., ``{Dark Energy Survey Year 3
  results: cosmology with moments of weak lensing mass maps \textendash{}
  validation on simulations}'', {\em Mon. Not. Roy. Astron. Soc.} {\bfseries
  498} (2020), no.~3, 4060--4087,
  \href{https://arxiv.org/abs/1911.05568}{{\ttfamily [arXiv:1911.05568]}}.

\bibitem[{Gilman} et~al.(2020){Gilman}, {Birrer}, and
  {Treu}]{2020A&A...642A.194G}
{Gilman}, D., {Birrer}, S., and {Treu}, T., ``{TDCOSMO. III. Dark matter
  substructure meets dark energy. The effects of (sub)halos on strong-lensing
  measurements of H$_{0}$}'', {\em \aap} {\bfseries 642} (2020) A194,
  \href{https://arxiv.org/abs/2007.01308}{{\ttfamily [arXiv:2007.01308]}}.

\bibitem[Harvey et~al.(2020)Harvey, Valkenburg, Tamone, Boyarsky, Courbin, and
  Lovell]{Harvey:2019bco}
Harvey, D., Valkenburg, W., Tamone, A., Boyarsky, A., Courbin, F., and Lovell,
  M., ``{Exploiting flux ratio anomalies to probe warm dark matter in future
  large scale surveys}'', {\em Mon. Not. Roy. Astron. Soc.} {\bfseries 491}
  (2020), no.~3, 4247--4253,
  \href{https://arxiv.org/abs/1912.02196}{{\ttfamily [arXiv:1912.02196]}}.

\bibitem[Kainulainen and Marra(2009)]{Kainulainen:2009dw}
Kainulainen, K. and Marra, V., ``{A new stochastic approach to cumulative weak
  lensing}'', {\em Phys. Rev. D} {\bfseries 80} (2009) 123020,
  \href{https://arxiv.org/abs/0909.0822}{{\ttfamily [arXiv:0909.0822]}}.

\bibitem[{Kantowski}(1969)]{1969ApJ...155...89K}
{Kantowski}, R., ``{Corrections in the Luminosity-Redshift Relations of the
  Homogeneous Friedmann Models}'', {\em \apj} {\bfseries 155} (1969) 89.

\bibitem[{Keeton} et~al.(1997){Keeton}, {Kochanek}, and
  {Seljak}]{1997ApJ...482..604K}
{Keeton}, C.~R., {Kochanek}, C.~S., and {Seljak}, U., ``{Shear and Ellipticity
  in Gravitational Lenses}'', {\em \apj} {\bfseries 482} (1997), no.~2,
  604--620,  \href{https://arxiv.org/abs/astro-ph/9610163}{{\ttfamily
  [arXiv:astro-ph/9610163]}}.

\bibitem[{Kovner}(1987)]{1987ApJ...316...52K}
{Kovner}, I., ``{The Thick Gravitational Lens: A Lens Composed of Many Elements
  at Different Distances}'', {\em \apj} {\bfseries 316} (1987) 52.

\bibitem[Kuhn et~al.(2020)Kuhn, Bruderer, Birrer, Amara, and
  R\'efr\'egier]{Kuhn:2020wpy}
Kuhn, F.~A., Bruderer, C., Birrer, S., Amara, A., and R\'efr\'egier, A.,
  ``{Combining strong and weak lensing estimates in the Cosmos field}'',
  \href{https://arxiv.org/abs/2010.08680}{{\ttfamily [arXiv:2010.08680]}}.

\bibitem[Li et~al.(2020)Li, Becker, and Dye]{Li:2020fpq}
Li, N., Becker, C., and Dye, S., ``{The impact of line-of-sight structures on
  measuring $H_0$ with strong lensing time-delays}'',
  \href{https://arxiv.org/abs/2006.08540}{{\ttfamily [arXiv:2006.08540]}}.

\bibitem[McCully et~al.(2014)McCully, Keeton, Wong, and
  Zabludoff]{McCully:2013fga}
McCully, C., Keeton, C.~R., Wong, K.~C., and Zabludoff, A.~I., ``{A New Hybrid
  Framework to Efficiently Model Lines of Sight to Gravitational Lenses}'',
  {\em Mon. Not. Roy. Astron. Soc.} {\bfseries 443} (2014), no.~4, 3631--3642,
  \href{https://arxiv.org/abs/1401.0197}{{\ttfamily [arXiv:1401.0197]}}.

\bibitem[McCully et~al.(2017)McCully, Keeton, Wong, and
  Zabludoff]{McCully:2016yfe}
McCully, C., Keeton, C.~R., Wong, K.~C., and Zabludoff, A.~I., ``{Quantifying
  Environmental and Line-of-Sight Effects in Models of Strong Gravitational
  Lens Systems}'', {\em Astrophys. J.} {\bfseries 836} (2017), no.~1, 141,
  \href{https://arxiv.org/abs/1601.05417}{{\ttfamily [arXiv:1601.05417]}}.

\bibitem[Migkas et~al.(2020)Migkas, Schellenberger, Reiprich, Pacaud,
  Ramos-Ceja, and Lovisari]{Migkas:2020fza}
Migkas, K., Schellenberger, G., Reiprich, T., Pacaud, F., Ramos-Ceja, M., and
  Lovisari, L., ``{Probing cosmic isotropy with a new X-ray galaxy cluster
  sample through the $L_{\text{X}}-T$ scaling relation}'', {\em Astron.
  Astrophys.} {\bfseries 636} (2020) A15,
  \href{https://arxiv.org/abs/2004.03305}{{\ttfamily [arXiv:2004.03305]}}.

\bibitem[{Petters} et~al.(2001){Petters}, {Levine}, and
  {Wambsganss}]{2001stgl.book.....P}
{Petters}, A.~O., {Levine}, H., and {Wambsganss}, J., ``{Singularity theory and
  gravitational lensing}'', 2001.

\bibitem[{Refsdal}(1964)]{1964MNRAS.128..307R}
{Refsdal}, S., ``{On the possibility of determining Hubble's parameter and the
  masses of galaxies from the gravitational lens effect}'', {\em \mnras}
  {\bfseries 128} (1964) 307.

\bibitem[{Sachs}(1961)]{1961RSPSA.264..309S}
{Sachs}, R., ``{Gravitational Waves in General Relativity. VI. The Outgoing
  Radiation Condition}'', {\em Proceedings of the Royal Society of London
  Series A} {\bfseries 264} (1961), no.~1318, 309--338.

\bibitem[Saunders(1969)]{saunders_observations_1969}
Saunders, P.~T., ``Observations in some simple cosmological models with
  shear'', {\em Month. Not. R. Astron. Soc.} {\bfseries 142} (1969) 213.

\bibitem[Schneider(1997)]{Schneider:1997bq}
Schneider, P., ``{The Cosmological lens equation and the equivalent single
  plane gravitational lens}'', {\em Mon. Not. Roy. Astron. Soc.} {\bfseries
  292} (1997) 673,
 \href{https://arxiv.org/abs/astro-ph/9706185}{{\ttfamily
  [arXiv:astro-ph/9706185]}}.

\bibitem[Schneider(2019)]{Schneider:2014vka}
Schneider, P., ``{Generalized multi-plane gravitational lensing: time delays,
  recursive lens equation, and the mass-sheet transformation}'', {\em Astron.
  Astrophys.} {\bfseries 624} (2019) A54,
  \href{https://arxiv.org/abs/1409.0015}{{\ttfamily [arXiv:1409.0015]}}.

\bibitem[{Schneider} et~al.(1992){Schneider}, {Ehlers}, and
  {Falco}]{1992grle.book.....S}
{Schneider}, P., {Ehlers}, J., and {Falco}, E.~E., ``{Gravitational Lenses}'',
  1992.

\bibitem[{Secrest} et~al.(2021){Secrest}, {von Hausegger}, {Rameez},
  {Mohayaee}, {Sarkar}, and {Colin}]{2021ApJ...908L..51S}
{Secrest}, N.~J., {von Hausegger}, S., {Rameez}, M., {Mohayaee}, R., {Sarkar},
  S., and {Colin}, J., ``{A Test of the Cosmological Principle with Quasars}'',
  {\em \apjl} {\bfseries 908} (2021), no.~2, L51,
  \href{https://arxiv.org/abs/2009.14826}{{\ttfamily [arXiv:2009.14826]}}.

\bibitem[{Seitz} and {Schneider}(1994)]{1994A&A...287..349S}
{Seitz}, S. and {Schneider}, P., ``{Some remarks on multiple deflection
  gravitational lensing.}'', {\em \aap} {\bfseries 287} (1994) 349--360.

\bibitem[{Tihhonova} et~al.(2018){Tihhonova}, {Courbin}, {Harvey}, {Hilbert},
  {Rusu}, {Fassnacht}, {Bonvin}, {Marshall}, {Meylan}, {Sluse}, {Suyu}, {Treu},
  and {Wong}]{2018MNRAS.477.5657T}
{Tihhonova}, O., {Courbin}, F., {Harvey}, D., {Hilbert}, S., {Rusu}, C.~E.,
  {Fassnacht}, C.~D., {Bonvin}, V., {Marshall}, P.~J., {Meylan}, G., {Sluse},
  D., {Suyu}, S.~H., {Treu}, T., and {Wong}, K.~C., ``{H0LiCOW VIII. A
  weak-lensing measurement of the external convergence in the field of the
  lensed quasar HE 0435-1223}'', {\em \mnras} {\bfseries 477} (2018), no.~4,
  5657--5669,  \href{https://arxiv.org/abs/1711.08804}{{\ttfamily
  [arXiv:1711.08804]}}.

\bibitem[{Wong} et~al.(2020){Wong}, {Suyu}, {Chen}, {Rusu}, {Millon}, {Sluse},
  {Bonvin}, {Fassnacht}, {Taubenberger}, {Auger}, {Birrer}, {Chan}, {Courbin},
  {Hilbert}, {Tihhonova}, {Treu}, {Agnello}, {Ding}, {Jee}, {Komatsu},
  {Shajib}, {Sonnenfeld}, {Blandford}, {Koopmans}, {Marshall}, and
  {Meylan}]{2020MNRAS.498.1420W}
{Wong}, K.~C., {Suyu}, S.~H., {Chen}, G. C.~F., {Rusu}, C.~E., {Millon}, M.,
  {Sluse}, D., {Bonvin}, V., {Fassnacht}, C.~D., {Taubenberger}, S., {Auger},
  M.~W., {Birrer}, S., {Chan}, J. H.~H., {Courbin}, F., {Hilbert}, S.,
  {Tihhonova}, O., {Treu}, T., {Agnello}, A., {Ding}, X., {Jee}, I., {Komatsu},
  E., {Shajib}, A.~J., {Sonnenfeld}, A., {Blandford}, R.~D., {Koopmans}, L.
  V.~E., {Marshall}, P.~J., and {Meylan}, G., ``{H0LiCOW {\textendash} XIII. A
  2.4 per cent measurement of H$_{0}$ from lensed quasars:
  5.3{\ensuremath{\sigma}} tension between early- and late-Universe probes}'',
  {\em \mnras} {\bfseries 498} (2020), no.~1, 1420--1439,
  \href{https://arxiv.org/abs/1907.04869}{{\ttfamily [arXiv:1907.04869]}}.

\end{thebibliography}

\end{document}